\documentclass[journal,10pt]{IEEEtran}
\usepackage{algpseudocode}                                 
\usepackage{algorithm}
\usepackage{graphicx}                                      
\usepackage{amsmath}
\usepackage{amssymb}
\usepackage{amsfonts}
\usepackage{booktabs}
\usepackage{multirow}
\usepackage[subrefformat=parens,farskip=0pt,justification=centering]{subfig}
\usepackage{color}
\usepackage{cite}                                          
\usepackage{comment}                                       
\usepackage{soul}                                          
\soulregister\cite7
\soulregister\ref7
\soulregister\pageref7
\usepackage{amsthm}
\usepackage{etoolbox}                                      
\usepackage{url}
\usepackage{nth}                                           
\usepackage{bm}                                            
\usepackage{balance}
\usepackage{threeparttable}
\usepackage[bookmarks=false]{hyperref}                     %
\hypersetup{
    colorlinks = true,
    citecolor  = blue,
    linkcolor  = blue,
    urlcolor   = blue,
}
\usepackage{filecontents}                                  
\usepackage{pgfplots}
\usepackage{pgfplotstable}
\pgfplotsset{compat=newest}
\newtheorem{myproblem}{\textbf{Problem}}

\algrenewcommand\textproc{\texttt}

\makeatletter
\let\OldStatex\Statex
\renewcommand{\Statex}[1][3]{%
  \setlength\@tempdima{\algorithmicindent}%
  \OldStatex\hskip\dimexpr#1\@tempdima\relax
}
\makeatother

\RequirePackage[normalem]{ulem} 
\RequirePackage{color}\definecolor{RED}{rgb}{1,0,0}\definecolor{BLUE}{rgb}{0,0,1} 

\newcommand{\norm}[1]{\lVert#1\rVert}
\renewcommand{\vec}[1]{\boldsymbol{#1}}

\newcommand{\bigmax}[2]{\max\left({#1}, {#2}\right)}

\newcommand{\bigbracket}[1]{\left[{#1}\right]}

\newcommand{\eqRef}[1]{Eq.~\eqref{#1}}
\newcommand{\tabRef}[1]{TABLE~\ref{#1}}
\newcommand{\figRef}[1]{Fig.~\ref{#1}}
\newcommand{\secRef}[1]{Section~\ref{#1}}

\newcommand{\colorstuff}[2]{\color{#1}{#2}\color{black}}

\graphicspath{{../figs/}}

\begin{document}

\title{
Multi-Electrostatic FPGA Placement Considering SLICEL-SLICEM Heterogeneity, Clock Feasibility, and Timing Optimization
}

\author{
    Jing Mai,
    Jiarui Wang,
    Zhixiong Di \textit{Member, IEEE},
    Yibo Lin \textit{Member, IEEE}\\
\thanks{The preliminary version has been presented at the Design Automation Conference (DAC) in 2022.
This work was supported in part by the National Science Foundation of China (Grant No. 62034007 and No. 62141404) and the 111 Project (B18001).
}
\thanks{J.~Mai and J.~Wang are with School of Computer Science, Peking University, Beijing, China.}
\thanks{Y.~Lin is with School of Integrated Circuits, Peking University, Beijing, China. Corresponding author: Yibo Lin (yibolin@pku.edu.cn).}
\thanks{Z.~Di is with School of Information Science and Technology, Southwest Jiaotong University, Chengdu, China.}
}

\maketitle
\thispagestyle{empty} 

\begin{abstract}
When modern FPGA architecture becomes increasingly complicated, modern FPGA
placement is a mixed optimization problem with multiple objectives, including
wirelength, routability, timing closure, and clock feasibility.
Typical FPGA devices nowadays consist of heterogeneous SLICEs like SLICEL and
SLICEM. The resources of a SLICE can be configured to \{LUT, FF, distributed RAM,
SHIFT, CARRY\}.
Besides such heterogeneity, advanced FPGA architectures also bring complicated
constraints like timing, clock routing, carry chain alignment, etc. The above
heterogeneity and constraints impose increasing challenges to FPGA placement
algorithms. 

In this work, we propose a multi-electrostatic FPGA placer considering
the aforementioned SLICEL-SLICEM heterogeneity under timing, clock routing and
carry chain alignment constraints. We first propose an effective SLICEL-SLICEM
heterogeneity model with a novel electrostatic-based density formulation.
We also design a dynamically adjusted preconditioning and carry chain alignment
technique to stabilize the optimization convergence. 
We then propose a timing-driven net weighting scheme to incorporate timing
optimization.
%
Finally, we put forward a nested Lagrangian relaxation-based placement
framework to incorporate the optimization objectives of wirelength,
routability, timing, and clock feasibility. 
Experimental results on both academic and industrial benchmarks demonstrate
that our placer outperforms the state-of-the-art placers in quality and
efficiency. 
\end{abstract}

\section{Introduction}
\label{sec:Introduction}

Placement is a critical step in the FPGA design flow, with a great impact on
routability and timing closure.
In the literature, three types of FPGA placement have been investigated: 1)
partitioning-based, 2) simulated annealing (SA), and 3) analytical
approaches~\cite{PLACE_2008_Lee, PLACE_PIEEE2015_Markov}.
Partitioning-based approaches such as \cite{PLACE_TCAD2005_Maidee} usually have
good scalability, but often fail to achieve high-quality results.
%
SA-based approaches like the widely-adopted academic tool
\texttt{VPR}\cite{PLACE_TCAD2005_Maidee} can achieve good results on small
designs, but suffer from poor scalability on large designs.
Recent studies have shown that analytical approaches~\cite{PLACE_DAC2015_ShengYen,
PLACE_TCAD2018_Li, PLACE_ICCAD2016_Pui_RippleFPGA,
PLACE_ICCAD2016_Ryan,PLACE_TCAD2019_Li_UTPLACEF_DL,
PLACE_TODAES2018_Li_UTPlaceF2, PLACE_FPGA2019_Li,
PLACE_ICCAD2017_Pui_RippleFPGA, PLACE_ICCAD21_Liang_AMFPlacer,
PLACE_TCAD2021_Meng, PLACE_ICCAD2017_Kuo_NTUfplace, PLACE_TCAD2020_Chen}
can achieve the best trade-off between quality and runtime. 
Thus modern FPGA placers mainly adopt analytical approaches in academia and
industry.

Modern FPGA placement has two major challenges: 1) the heterogeneity of FPGA
architecture, 2) the various constraints (e.g., timing, clock routing, chain
alignment, etc.) imposed by advanced circuit
designs~\cite{BENCH_ISPD2016_PLACE, BENCH_ISPD2017_PLACE,
PLACE_ICCAD2013_Contest, PLACE_PIEEE2015_Markov}.
The heterogeneity of the FPGA architecture comes from the variety of instance
types, the imbalance of resource distribution, and the asymmetric slice
compatibility from the SLICEL-SLICEM heterogeneity~\cite{DI_ULTRASCALE_CLB}.
The diversity of instance sizes and the inconsecutive site compatibility challenge
the modern FPGA placement algorithms, which are mainly based on continuous
optimization \cite{PLACE_PIEEE2015_Markov, PLACE_TCAD2021_Meng}.

Furthermore, solving the highly heterogeneous FPGA placement problem while
satisfying advanced constraints has become more
challenging~\cite{PLACE_PIEEE2015_Markov}.
i)
Wire-induced delays are becoming the primary source of overall circuit delay
\cite{PLACE_DAC2019_Martin, PLACE_DATE2021_Lin, PLACE_FPGA2000_Marquardt}.
%
Timing-driven placement is required to meet the aggressive timing constraints.
%
%
However, the nonlinear and nonmonotonic wire delays
impose unique challenges to timing optimization in FPGA
placement.
ii) Modern FPGA adopts complicated clock architectures to achieve low clocking
skew and high performance~\cite{BENCH_ISPD2016_PLACE, BENCH_ISPD2017_PLACE}.
Such a clock architecture introduces complicated clock routing constraints,
increasing the challenges in FPGA placement.
Therefore, clock-aware FPGA placement is required to accommodate the needs of
modern FPGA design flows.
iii) Modern FPGA needs to align the cascaded instances like CARRY into an aligned chain at the placement stage to boost performance.
Such a chain alignment requirement induces large placement blocks and
tends to degrade the quality of the solution.

\begin{figure}[tb]
    \centering
    \subfloat[]{\includegraphics[height=0.4\columnwidth]{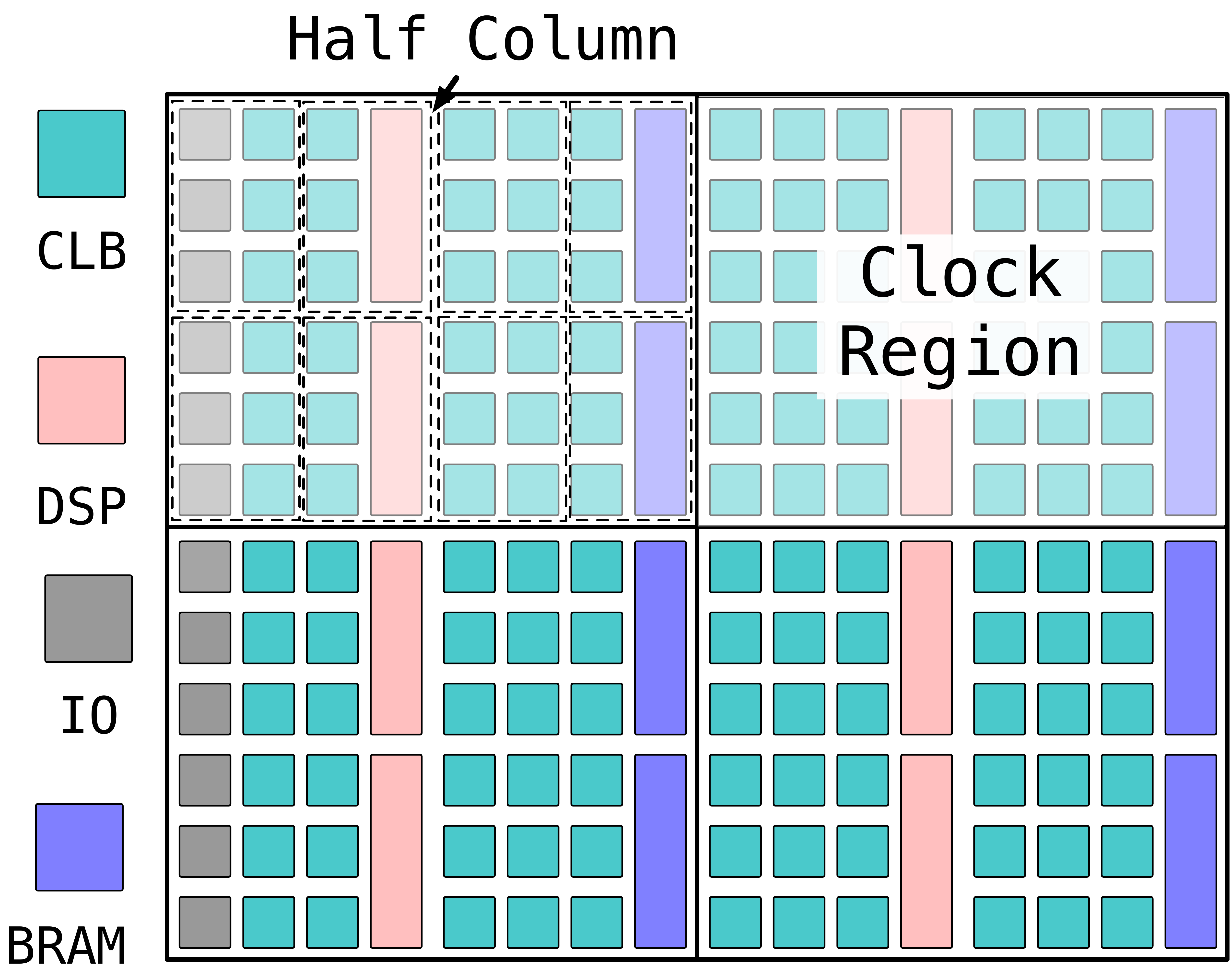}\label{fig:fpga_arch}} \hfill
    \subfloat[]{\includegraphics[height=0.4\columnwidth]{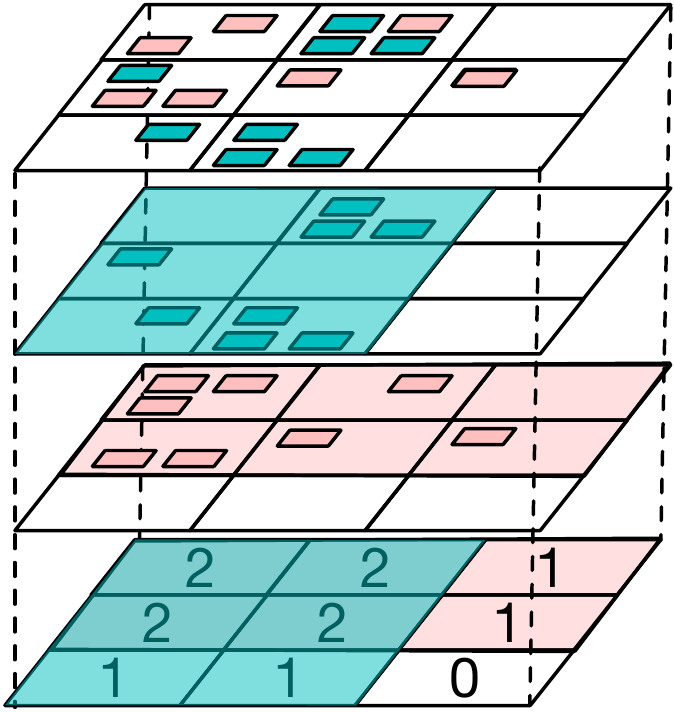}\label{fig:cr_con}} \\
    \vspace{-.1in}
    \subfloat[]{\includegraphics[width=0.75\columnwidth]{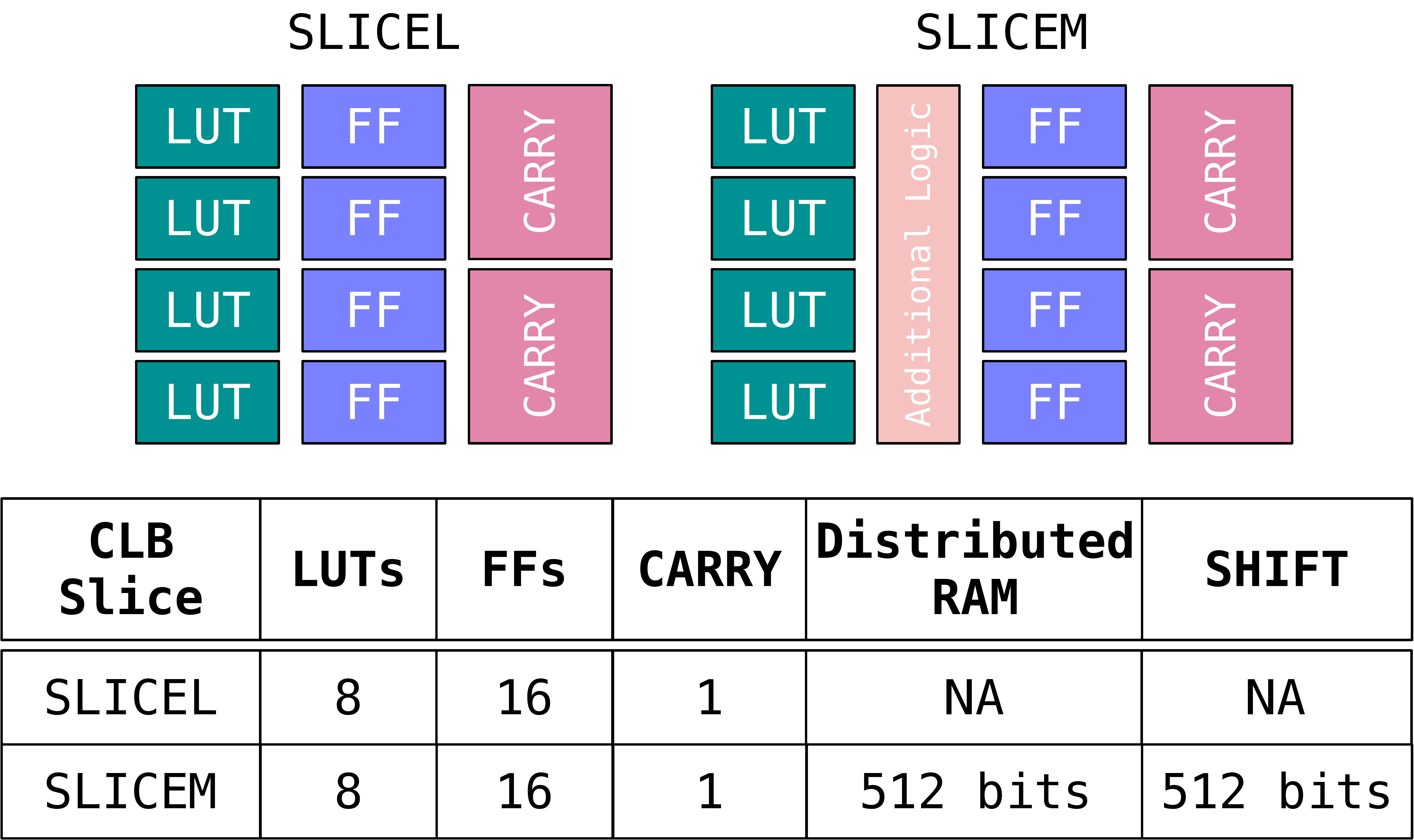}\label{fig:CLB_slice}}
    \caption{
      (a) An example of a simplified vertical FPGA architecture depiction showing a $2 \times 2$ clock region and half columns (dash lines) for \emph{Xilinx UltraScale}.
      (b) An example of how the clock demand can be calculated. The different colors represent different clocks, and the numbers represent the clock demand of each CR.
      (c) An illustration of CLB slices classified into SLICEL and SLICEM with asymmetric compatibility. 
        In a SLICEL, LUT blocks can be configured to be LUTs.
        A SLICEM can only be configured in \textit{one} of the following modes: LUT, distributed RAM, or SHIFT.     
        There is no mixing of LUTs, distributed RAMs, and SHIFTs in a CLB. 
    }
    \vspace{-.2in}
\end{figure}

In this work, we propose a state-of-the-art placement framework considering
SLICEL-SLCIEM heterogeneity and the co-optimation with wirelength, routability,
clock feasibility, and timing optimization.
We handle a comprehensive set of instance types, i.e, \{ LUT, FF, BRAM,
distributed RAM, SHIFT, CARRY \}, and cope with SLICEL-SLCIEM heterogeneity
based on a multi-electrostatic system.
We propose a uniform non-linear optimization paradigm taking wirelength,
routability, clock feasibility, and timing optimization into consideration from
the perspective of \textit{nested Lagrangian method}.
The main contributions of this work are summarized as follows.
\begin{itemize}
\item We adopt an effective SLICEL-SLICEM heterogeneity model based on the
division and assembly of electrostatic-based density formulation. 
\item We develop a dynamically adjusted preconditioning and carry chain
  alignment technique to stabilize the optimization convergence and enable better
    final placement results. 
\item We cope with the time violation by an effective timing-criticality-based
  net weighting scheme, and incorporate the timing optimization into a continuous
    optimization algorithm. 
\item To achieve effective clock routing violation elimination, we adopt a
  instance-to-clock-region mapping considering the resource capacity of the clock
    regions and perturbation to the placement, and propose a quadratic clock
    penalty function in a continuous global placement engine with minor quality
    degradation.
\item Putting the aforementioned techniques together, we put forward a nested
  Lagrangian relaxation framework incorporating the optimization objectives of
    wirelength, routability, timing, and clock feasibility.
\end{itemize}
Experiments on \textit{ISPD 2017 contest benchmarks} demonstrate 14.2\%,
11.7\%, 9.6\%, and 7.9\% improvement in routed wirelength, compared to the
recent cutting-edge FPGA placers~\cite{PLACE_TODAES2018_Li_UTPlaceF2,
PLACE_ICCAD2017_Pui_RippleFPGA, PLACE_FPGA2019_Li, PLACE_TCAD2020_Chen},
respectively. 
Our placer also supports GPU acceleration and gains 1.45-6.58$\times$ speedup over the baselines.
Further experiments on \textit{industrial} benchmarks demonstrate that the
proposed algorithms can achieve 23.6\% better WNS, 22.5\% better TNS with about 2\% routed wirelength degradation compared with the conference version.

The rest of the paper is organized as follows.
\secRef{sec:Preliminary} introduces the preliminary knowledge of the FPGA
architecture and modern FPGA placement.
\secRef{sec:Algorithm} details the core placement algorithms.
\secRef{sec:Results} shows the experimental results, followed by the conclusion
in \secRef{sec:Conclusion}.

\begin{table*}[tb]
\caption{Features of the published state-of-the-art FPGA placers.}
\label{tab:FPGAPlacers}
  \resizebox{1.0\textwidth}{!}{
\begin{tabular}{|cc|cccccc|ccccc|}
\hline
\multicolumn{2}{|c|}{Placer}                                                                                                                                                    
  & \begin{tabular}[c]{@{}c@{}}\texttt{RippleFPGA}\\ \cite{TCAD18_RippleFPGA_Chen}\end{tabular}
  & \begin{tabular}[c]{@{}c@{}}\texttt{GPlace}\\ \cite{PLACE_ICCAD2016_Ryan}\end{tabular}
  & \begin{tabular}[c]{@{}c@{}}\texttt{UTPlaceF}\\ \cite{PLACE_TCAD2018_Li}\end{tabular}
  & \begin{tabular}[c]{@{}c@{}}\texttt{elfPlace}\\ \cite{PLACE_TCAD2021_Meng}\end{tabular}
  & \begin{tabular}[c]{@{}c@{}}\texttt{FTPlace}\\ \cite{PLACE_DAC2019_Martin}\end{tabular}
  & \begin{tabular}[c]{@{}c@{}}\texttt{GPlace}\\ \texttt{3.0} \cite{TODAES18_GPlace3_Abuowaimer}\end{tabular}
  & \begin{tabular}[c]{@{}c@{}}\texttt{RippleFPGA}\\ \texttt{Clock-Aware} \cite{PLACE_ICCAD2017_Pui_RippleFPGA}\end{tabular}
  & \begin{tabular}[c]{@{}c@{}}\texttt{UTPlaceF}\\ \texttt{{2.0}}\&\texttt{2.X} \cite{PLACE_TODAES2018_Li_UTPlaceF2, PLACE_FPGA2019_Li}\end{tabular}
  & \begin{tabular}[c]{@{}c@{}}\texttt{NTUfPlace}\\ \cite{PLACE_TCAD2020_Chen}\end{tabular}
  & \begin{tabular}[c]{@{}c@{}}Lin \textit{et al.}\\ \cite{PLACE_DATE2021_Lin}\end{tabular}
  & Ours         \\ \hline
\multicolumn{2}{|c|}{Clock Constraints}                                                                                                                                         & $\times$     & $\times$     & $\times$     & $\times$     & $\times$     & $\times$                                              & $\checkmark$                                                     & $\checkmark$                                                & $\checkmark$ & $\checkmark$ & $\checkmark$ \\ \hline
\multicolumn{1}{|c|}{\multirow{2}{*}{\begin{tabular}[c]{@{}c@{}}Resources\\ Supported\end{tabular}}} & \begin{tabular}[c]{@{}c@{}}LUT, FF, \\ BRAM, DSP\end{tabular}            & $\checkmark$ & $\checkmark$ & $\checkmark$ & $\checkmark$ & $\checkmark$ & $\checkmark$                                         & $\checkmark$                                                     & $\checkmark$                                                & $\checkmark$ & $\checkmark$ & $\checkmark$ \\ \cline{2-13} 
\multicolumn{1}{|c|}{}                                                                               & \begin{tabular}[c]{@{}c@{}}CARRY, SHIFT, \\ Distributed RAM\end{tabular} & $\times$     & $\times$     & $\times$     & $\times$     & $\times$     & $\times$                                             & $\times$                                                         & $\times$                                                    & $\times$     & $\times$     & $\checkmark$ \\ \hline
\multicolumn{2}{|c|}{Timing Optimization}                                                                                                                                       & $\times$     & $\times$     & $\times$     & $\times$     & $\checkmark$ & $\times$                                             & $\times$                                                         & $\times$                                                    & $\times$     & $\checkmark$ & $\checkmark$ \\ \hline
\multicolumn{2}{|c|}{GPU-Acceleration}                                                                                                                                          & $\times$     & $\times$     & $\times$     & $\checkmark$ & $\times$     & $\times$                                             & $\times$                                                         & $\times$                                                    & $\times$     & $\times$     & $\checkmark$ \\ \hline
\multicolumn{2}{|c|}{Algorithm Category}                                                                                                                                        & Quadratic    & Quadratic    & Quadratic    & Nolinear     & Quadratic    & Quadratic                                            & Quadratic                                                        & Quadratic                                                   & Nonlinear    & Nonlinear    & Nonlinear    \\ \hline
\end{tabular}}
\end{table*}

\section{Preliminaries}
\label{sec:Preliminary}

In this section, we primarily focus on the architecture of FPGAs and the methodology of multi-electrostatics-based FPGA placement.


\subsection{Device Architecture}
In this work, we use the \textit{Xilinx UltraScale} family~\cite{DI_ULTRASCALE, DI_ULTRASCALE_CLB},e.g., the \textit{UltraScale VU095} as a model FPGA design (illustrated in \figRef{fig:fpga_arch}).
The ISPD 2016 and 2017 FPGA placement challenges employed a condensed version of this architecture with a limited number of instance types, including LUT, FF, BRAM, and DSP~\cite{BENCH_ISPD2016_PLACE, BENCH_ISPD2017_PLACE}. 

\subsubsection{SLICEL-SLICEM heterogeneity}
\label{sec:clb_heterogeneity}
Shift registers (SHIFT) and distributed RAMs are two additional LUT-like instances of the architecture in addition to regular LUTs. 
Slices in CLBs fall into two categories: SLICEL and SLICEM, whose architectures are depicted in \figRef{fig:CLB_slice}. 
Due to the modest differences in logic resources, SLICEL and SLICEM support various configurations.
LUTs, FFs, and CARRYs can be placed in both SLICEL and SLICEM. 
On the other hand, SHIFT and distributed RAMs can only be placed in SLICEM, which sets them apart from other common instances such as DSPs and BRAMs. 
Additionally, a SLICEM cannot be used as SHIFTs or distributed RAMs if it is configured as LUTs; vice versa. 

\subsubsection{Carry Chain Alignment Constraints}
As a placement constraint for CARRY instances, we also need to take carry chain alignment into account. 
A carry chain is made up of several consecutive CARRY instances connected by cascaded wires from the lower bits to the upper bits, 
and the CARRY instances are arranged in CLB slices.
According to the alignment constraint, each CARRY instance in a chain must be positioned in a single column and in subsequent slices in the correct sequence for the cascading wires.


\subsubsection{Clock Constraints}
\label{sec:clock_constraint}

The target FPGA device owns $5 \times 8$ rectangular-shaped clock regions (CRs) in a grid manner, as shown in \figRef{fig:fpga_arch}. 
\footnote{
\figRef{fig:fpga_arch} only contains part of the whole $5 \times 8$ CRs, but is sufficient for illustration.
}.
Each CR is made up of columns of site resources and can be further horizontally subdivided into pairs of lower and upper half columns (HCs) of half-clock-region height. 
Except for a few corner cases, the width of each HC is the same as that of two site columns, as shown in \figRef{fig:fpga_arch}.
All the clock sinks within a half column are driven by the same leaf clock tracks.

The clock routing architecture imposes two clock constraints on placement, i.e., the \emph{clock region constraint} and the \emph{half column constraint}, as shown in \figRef{fig:cr_con}.
The clock region constraint limits each clock region's clock demand to a maximum  of 24 clock nets, where the clock demand is the total number of clock nets whose bounding boxes intersect with the clock region. 
The half-column constraint limits the number of clock nets within the half-column to a maximum of 12.

\subsection{Multi-Electrostatics based FPGA Placement}
\label{sec:multi_electrostatics_placement}

\begin{figure}[h]
 \vspace{-.2in}
  \includegraphics[width=.4\textwidth]{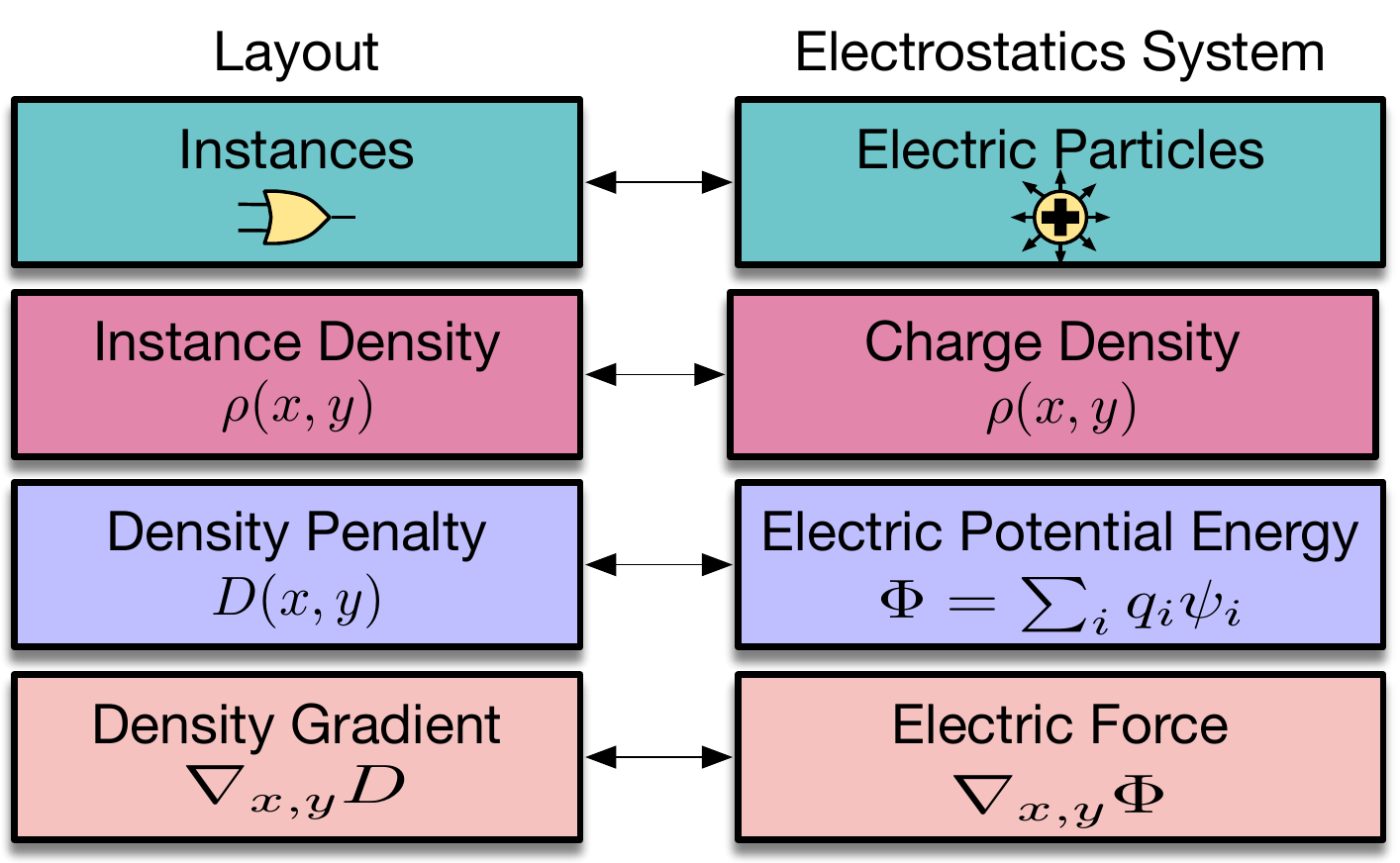}
  \caption{
    Analogy between placement for a single resource type and an electrostatic system \cite{PLACE_TCAD2015_Lu}. 
  }
  \label{fig:electrostatics_analogy}
\vspace{-.2in}
\end{figure}
\begin{figure*}[tb]
  \includegraphics[width=.98\textwidth]{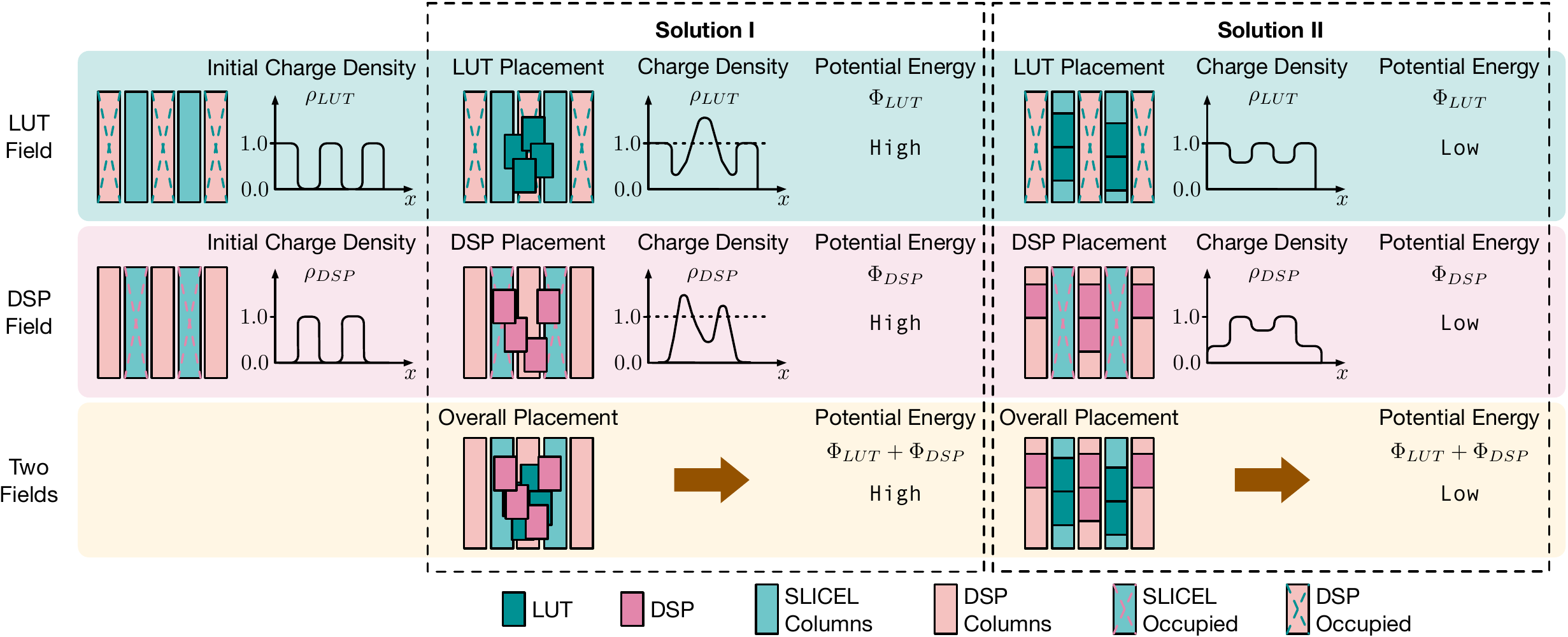}
  \caption{
  An example of a multi-electrostatic formulation for LUT and DSP resources,
  which correspond to two electric fields. 
  Unavailable columns are treated as occupied when calculating the initial charge density for each field. 
  Take DSP as an example. 
  Density overflow can occur if a DSP instance is not put in a DSP column or if there are overlaps between DSP instances, 
  resulting in an uneven density distribution of the field and, finally, excessive electric
  potential energy. 
  As a result, limiting energy in the layout can assist in the resolution of density overflow and spread cases.
  If we face density underflow ($<1.0$), we can insert fillers with positive charges for each field to fill the vacant spaces, \cite{PLACE_TODAES2015_Lu, PLACE_TCAD2021_Meng}.
  As a result, only density overflow will provide considerable potential energy.
  We may handle them in the same way by including FF and BRAM fields.
  }
  \label{fig:multi_electrostatics_exp}
  \vspace{-.1in}
\end{figure*}

As shown in \figRef{fig:electrostatics_analogy}, electrostatics-based placement models each instance as an electric particle in an electrostatic system. 
As firstly stated in the ASIC placement \cite{PLACE_TODAES2015_Lu}, \emph{minimizing potential energy} can
resolve density overflow in the layout.
The principle is based on the fundamental physical insight that a balanced charge
distribution in an electrostatic system contributes to low potential energy, so
\emph{minimizing potential energy can resolve density overflow and help spread
instances in the layout}. 
We are also extending this approach to the use of multiple electrostatic fields, which will enable multiple types of resource to be handled in FPGA placement, such as LUTs, FFs, DSPs, and BRAMs.
\figRef{fig:multi_electrostatics_exp} illustrates a multi-electrostatic formulation of LUTs and DSPs. 
In order to reduce density overflow, we must minimize the total potential energy of multiple fields because low energy means a balanced distribution of instances. 
The issue can be summarized as follows.
\begin{equation}
    \min_{\boldsymbol{x}, \boldsymbol{y}} \widetilde{W}(\boldsymbol{x}, \boldsymbol{y}) \quad \text { s.t. } \Phi_s(\boldsymbol{x}, \boldsymbol{y}) = 0,
    \label{eq:elfplace}
\end{equation}
where $\widetilde{W}(\cdot)$ is the wirelength objective, 
$\boldsymbol{x}, \boldsymbol{y}$ are instance locations, 
$S$ denotes the field type set, 
and $\Phi_s(\cdot)$ is the electric potential energy for field type $s \in S$.
Formally, we constrain the target potential energy of each field type to be zero, though the energy is usually non-negative.
The constraints can be further relaxed to the objective and guide the instances to spread out. 
Practically, we stop the optimization when the energy is small enough; or equivalently, the density overflow is low enough.
Notice that the formulation in \figRef{fig:multi_electrostatics_exp} assumes that one instance occupies the resources of only one field, 
which cannot handle the complicated SLICEL-SLICEM heterogeneity shown in \secRef{sec:clb_heterogeneity}. 

\subsection{Timing Optimization}
\label{sec:timing_optimization}
Timing-driven placement imposes more concern about timing closure than the total wirelength objective in wirelength-driven placement.
Worst negative slack (WNS) and total negative slack (TNS) are two widely adopted timing metrics. 
WNS is the maximum negative slack among all timing paths in the design, and TNS is the sum of all negative slacks of timing endpoints.
Thus, WNS and TNS are used to evaluate the timing performance of a design from the worst and the global view respectively,  and the smaller the WNS and TNS are, the worse the timing performance is. 
The timing-driven placement problem can be formulated as follows. 
 \begin{subequations}
   \label{eq:timing_placement}
  \begin{align}
    \min_{\boldsymbol{x}, \boldsymbol{y}} & \quad \mathcal{T}(\boldsymbol{x}, \boldsymbol{y}), \\
    \text{s.t.} & \quad \rho_s(\boldsymbol{x}, \boldsymbol{y}) \leq \hat{\rho}_s,\quad \forall s \in S,
  \end{align}
\end{subequations}
where $S$ is the instance type set, $\rho_s(\cdot)$ denotes the density for
instance type $s \in S$, and  $\hat{\rho}_s$ represents the target density for
instance type $s \in S$. 
The objective function $\mathcal{T}(\cdot)$ can be WNS, TNS, or the weighted sum of both.
Improving TNS requires collaborative optimization of all timing paths, and is
therefore suitable for the global placement stage. 
On the other hand, WNS is more suitable for the detailed placement stage, as it
only considers the worst timing path. 
It is worth noting that directly solving \eqRef{eq:timing_placement} is very
difficult, because the delay model generally has strong discrete and non-convex
properties \cite{PLACE_ISPD2002_Kahng}.
Therefore, we draw on the two intuitive elements of wirelength-driven
placement and static timing analysis, i.e., net weights and slacks, to tackle
this problem. 

\subsection{Problem Formulation}
\tabRef{tab:FPGAPlacers} summarizes the characteristics of the published
state-of-the-art FPGA placers. In recent years, modern FPGA placers mainly
resort to quadratic programming-based approaches~\cite{ PLACE_DAC2015_ShengYen,
PLACE_TCAD2018_Li, PLACE_ICCAD2016_Pui_RippleFPGA, PLACE_ICCAD2016_Ryan,
PLACE_TCAD2019_Li_UTPLACEF_DL, PLACE_TODAES2018_Li_UTPlaceF2,
PLACE_FPGA2019_Li, PLACE_ICCAD2017_Pui_RippleFPGA,
PLACE_ICCAD21_Liang_AMFPlacer} and nonlinear optimization-based
approaches~\cite{ PLACE_TCAD2021_Meng, PLACE_ICCAD2017_Kuo_NTUfplace,
PLACE_TCAD2020_Chen} for the best trade-off between quality and efficiency.
Among them, the current state-of-the-art quality is achieved by nonlinear approaches \texttt{elfPlace}~\cite{PLACE_TCAD2021_Meng} 
and \texttt{NTUfPlace}~\cite{PLACE_TCAD2020_Chen}, whose instance density models are derived from a multi-electrostatics system and 
a hand-crafted bell-shaped field system. 
However, most existing FPGA placers only consider a simplified FPGA architecture, i.e., LUT, FF, DSP, and BRAM,
ignoring the commonplace SLICEL-SLCIEM heterogeneity in real FPGA architectures~\cite{PLACE_TCAD2018_Li,
PLACE_ICCAD2016_Pui_RippleFPGA,
PLACE_ICCAD2016_Ryan,PLACE_TCAD2019_Li_UTPLACEF_DL, PLACE_TCAD2021_Meng,
PLACE_TODAES2018_Li_UTPlaceF2, PLACE_FPGA2019_Li,
PLACE_ICCAD2017_Pui_RippleFPGA, PLACE_ICCAD2017_Kuo_NTUfplace,
PLACE_TCAD2020_Chen}. 
Among these placers, only a few placers utilize the parallelism that the GPU provides \cite{PLACE_TCAD2021_Meng},
and few consider timing and clock feasibility in practice \cite{
 PLACE_ICCAD2017_Pui_RippleFPGA, PLACE_ICCAD2017_Kuo_NTUfplace, PLACE_TCAD2020_Chen, PLACE_TCAD2019_Li_UTPLACEF_DL, PLACE_TODAES2018_Li_UTPlaceF2
}.

In this work, we aim at optimizing wirelength, timing, and routability while cooperating with SLICEL-SLICEM heterogeneity, alignment feasibility, and clock constraints.
We define the FPGA placement problem as follows.

\begin{myproblem}[FPGA Placement]

  \textit{Taking as input a netlist consisting of LUTs, FFs, DSPs, BRAMs,
distributed RAMs, SHIFTs, and CARRYs, generate a plausible FPGA placement
solution with optimized wirelength, timings, and routings, satisfying the
  requirements for alignment feasibility and meeting the clock constraints.}
\end{myproblem}

\section{Algorithms}
\label{sec:Algorithm}
We will detail the placement algorithm in this section.

\subsection{Overview of the Proposed Algorithm}
\label{sec:OverviewOfTheProposedFlow}

\begin{figure}[tb]
   \centering
   \includegraphics[width=\linewidth]{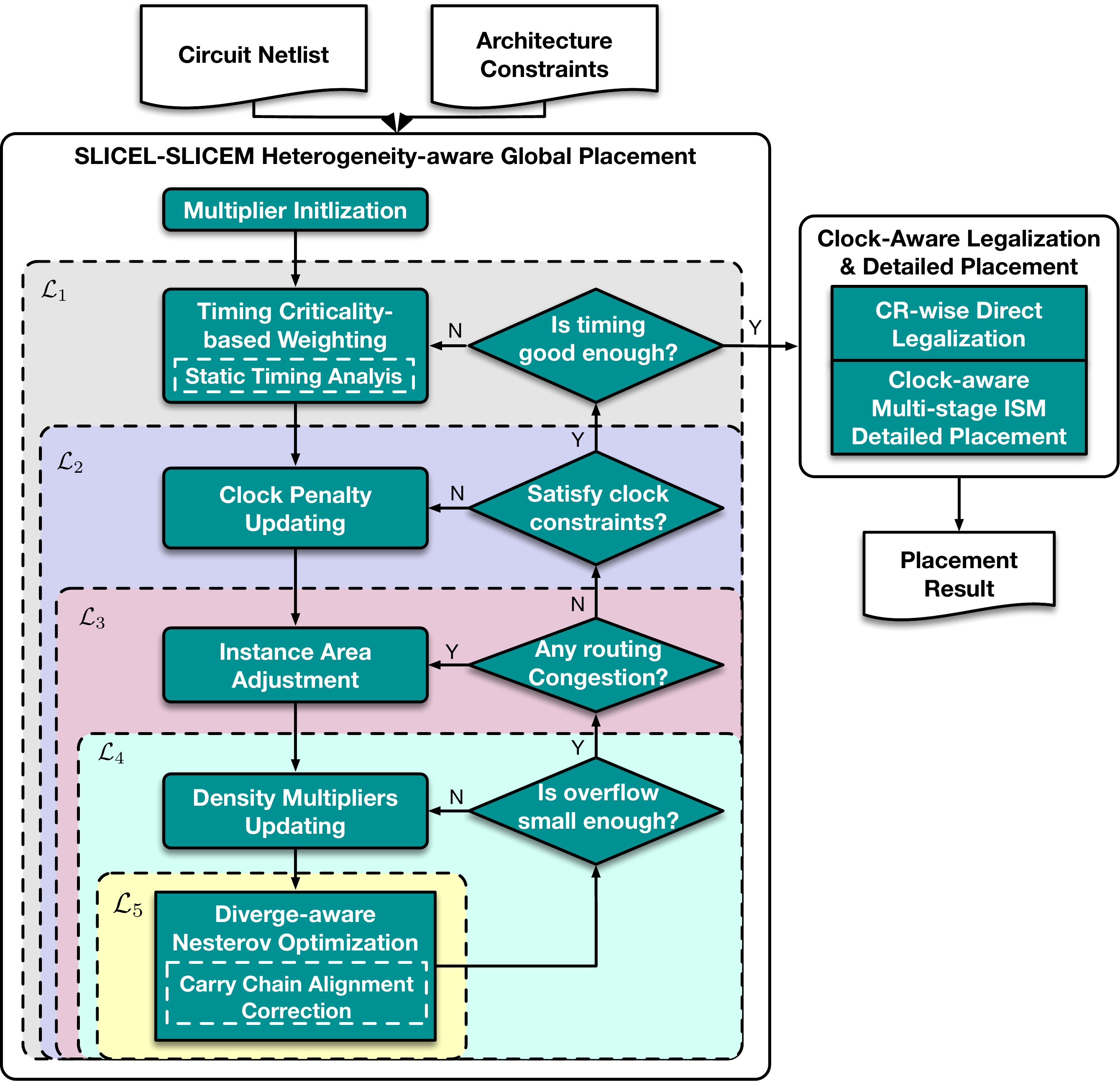}
   \caption{The proposed Overall Flow.}
   \label{fig:overall_flow}
   \vspace{-.2in}
\end{figure}

As illustrated in \figRef{fig:overall_flow}, our method includes two fundamental stages: (1)
nested global placement with timing awareness and clock feasibility, and (2)
clock-aware legalization and detailed placement.

We cope with the SLICEL-SLICEM heterogeneity by defining the field type set as $S = $ \{LUTL, LUTM-AL, FF, CARRY, DSP, BRAM\} with a special field setup (\secRef{sec:clb_heter}).
With clock constraints, carry chain alignment feasibility, and timing optimization, we formulate the problem as Formulation~(\ref{eq:opt_relax}).
 \begin{subequations}
   \label{eq:opt_relax}
  \begin{align}
    \min_{\boldsymbol{x}, \boldsymbol{y}} & \quad \widetilde{\mathcal{T}}_{\boldsymbol{\omega}}(\boldsymbol{x}, \boldsymbol{y}), \\
    \text{s.t.} & \quad \Phi_s(\boldsymbol{x}, \boldsymbol{y}; \mathcal{A}^s) = 0,\quad \forall s \in S, \\
    & \quad \varGamma(\boldsymbol{x}, \boldsymbol{y}) = 0, \\
    & \quad \text{\emph{Carry chain alignment constraint,}}
  \end{align}
\end{subequations}
$\widetilde{\mathcal{T}}_{\boldsymbol{\omega}}(\cdot)$ is the timing
performance objective,
where $\boldsymbol{\omega}$ measures the net criticality in the current timing graph
(\secRef{sec:timing_opt}).
$\mathcal{A}^s$ denotes the instance areas in the field $s$, 
and $\varGamma(\cdot)$ is the clock penalty term (\secRef{sec:cnp_algo}).
%
For brevity, in later discussions, we condense $\Phi_s(\boldsymbol{x}, \boldsymbol{y};\mathcal{A}^s)$ to $\Phi_s$ for all $s \in S$,
and denote $\boldsymbol{\Phi}$ as the potential energy vector, whose components are the potential energy for each field, i.e., $\Phi_s(\forall s \in S)$.

We relax the original problem (\ref{eq:opt_relax}) by leveraging the \emph{augmented Lagrangian method (ALM)} \cite{andreani2008augmented} to formulate a better unconstrained subproblem,
\begin{subequations}
\begin{align}
  \min_{\boldsymbol{x}, \boldsymbol{y}} \quad \mathcal{L}(\boldsymbol{x}, \boldsymbol{y}; \boldsymbol{\lambda}, \boldsymbol{\mathcal{A}},\eta, \boldsymbol{\omega}) & =
   \widetilde{\mathcal{T}}_{\boldsymbol{\omega}}(\boldsymbol{x}, \boldsymbol{y})  + \sum\limits_{s \in S} \lambda_s \mathcal{D}_s \nonumber \\
   & + \eta \varGamma(\boldsymbol{x}, \boldsymbol{y}), \\
\mathcal{D}_s &= \Phi_s + \frac{1}{2}\mathcal{C}_s \Phi_s^2,\quad \forall s \in S,
\end{align}
\label{eq:opt_alm}
\end{subequations}
The density multiplier vector is $\boldsymbol{\lambda} \in \mathbb{R}^{\vert S \vert}$,
and the clock penalty multiplier is $\eta \in \mathbb{R}$.
The purpose of the weighting coefficient vector $\boldsymbol{\mathcal{C}} \in \mathbb{R}^{\vert S \vert}$ is to achieve a balance between the first-order and second-order terms for density penalty.
We follow the setup for $\boldsymbol{\lambda}$ and $\boldsymbol{\mathcal{C}}$ as \cite{PLACE_TCAD2021_Meng}.
To cope with multiple constraints, we rewrite the problem in a nested manner through the Lagrangian relaxation method,
\begin{subequations}
  \begin{align}
    \textrm{Timing Opt.: } \quad \mathcal{L}_1 & = \max_{\boldsymbol{\omega}} \mathcal{L}_2 (\boldsymbol{\omega}),  \\
    \textrm{Clock Opt.: } \quad \mathcal{L}_2(\boldsymbol{\omega}) & = \max_{\eta} \mathcal{L}_3 (\eta, \boldsymbol{\omega}), \\
    \textrm{Routability Opt.: } \quad \mathcal{L}_3(\eta, \boldsymbol{\omega}) & = \max_{\boldsymbol{\mathcal{A}}} \mathcal{L}_4 (\boldsymbol{\mathcal{A}}, \eta, \boldsymbol{\omega}),  \\
    \textrm{Wirelength Opt.: } \quad \mathcal{L}_4(\boldsymbol{\mathcal{A}}, \eta, \boldsymbol{\omega}) & = \max_{\boldsymbol{\lambda}} \mathcal{L}_5 ( \boldsymbol{\lambda},  \boldsymbol{\mathcal{A}}, \eta, \boldsymbol{\omega}),\label{eq:nested_opt_density} \\
    \textrm{Subproblem: } \quad \mathcal{L}_5(\boldsymbol{\lambda}, \boldsymbol{\mathcal{A}}, \eta, \boldsymbol{\omega}) & = \min_{\boldsymbol{x}, \boldsymbol{y}} \mathcal{L}(\boldsymbol{x}, \boldsymbol{y}; \boldsymbol{\lambda}, \boldsymbol{\mathcal{A}},\eta, \boldsymbol{\omega}),
  \end{align}
  \label{eq:nested_opt}
\end{subequations}
where $\mathcal{L}_5$ denotes \eqRef{eq:opt_alm}.
To put it simply, a set of variables in the objective function is constrained to be the optimal solution of the extra optimization problem,
and the variables of the exterior optimization problem are passed toward the sub-problem as fixed parameters.
To illustrate, take the solving process of $\mathcal{L}_5$ as an example.
The subproblem $\mathcal{L}_5$ aims at finding the optimal instance positions $\boldsymbol{x}$ and $\boldsymbol{y}$
given a set of fixed parameters $\boldsymbol{\lambda}$, $\boldsymbol{\mathcal{A}}$, $\eta$, and $\boldsymbol{\omega}$ from $\mathcal{L}_4$.
When $\mathcal{L}_5$ comes to the convergence to its minimum solution,
the $\mathcal{L}_4$ optimizer will improve the parameter $\boldsymbol{\lambda}$
so as to enlarge the magnitude of the density terms in the overall optimization objectives.
Therefore, the $\mathcal{L}_5$ optimizer will try to find a better solution by moving the instances to a less dense region in a new iteration,
and thus the density constraints will be forced to be gradually adequate.
We also adopt the same procedure to solve the other subproblems.

\figRef{fig:overall_flow} depicts the nested loops to solve the problem.
$\mathcal{L}_1$ aims at improving the timing slacks via the timing-criticality-based weighting method.
The stopping criterion of $\mathcal{L}_1$ is whether timing slacks can be improved (\secRef{sec:timing_opt}).
$\mathcal{L}_2$ (\secRef{sec:cnp_algo}) aims at excluding the cases where clock routing is illegal under an analytical formulation.
We regard $\mathcal{L}_2$ as converged when there is no clock violation (\secRef{sec:clock_constraint}).
$\mathcal{L}_3$ develops the area inflation-based technique from \cite{PLACE_TCAD2021_Meng} in order to optimize the routability,
and the estimated routing congestion and pin density are the convergence criteria of $\mathcal{L}_3$.
$\mathcal{L}_4$ is the core wirelength-driven placement problem.
We empirically find that the density \emph{overflow} is a good indicator of the density constraints.
\footnote{
  We set the overflow threshold for LUTs and FFs to 10\% in our experiments.
}
For $\mathcal{L}_5$, we always solve with a fixed number of iterations, e.g., one iteration in the experiments.

In each iteration, we resolve the carry chain alignment constraints through
iterative support from \emph{iterative carry chain alignment correction}
(\secRef{sec:carry_alignment}).
Following placement, we evolve clock-aware direct legalization and detailed placement algorithms that above clock feasibility constraints are met \cite{PLACE_TCAD2019_Li_UTPLACEF_DL}.
We do not include details on routability optimization, clock-aware legalization, or detailed placement for brevity.

\subsection{Multi-Electrostatic Model for SLICEL-SLICEM Heterogeneity}
\label{sec:clb_heter}
%

\begin{figure}[tb]
  \centering
  \includegraphics[width=.48\textwidth]{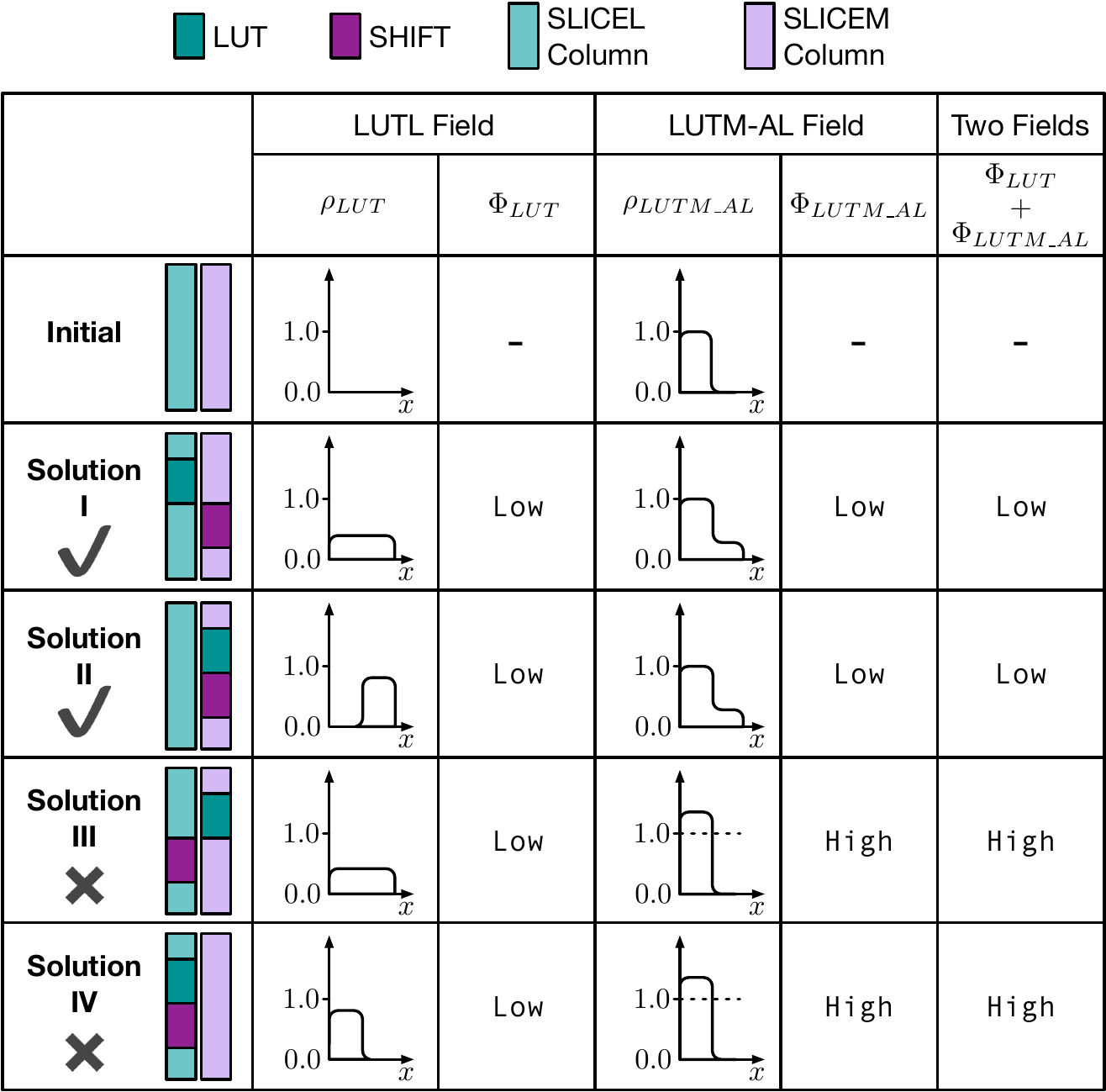}
   \caption{
     As an example, here is a special electrostatic field setup that handles
     asymmetric slice compatibility as a result of SLICEL-SLICEM heterogeneity.
     LUTM-AL fields are prone to overflowing density due to the SHIFT instance
     placed on the SLICEL column in Solution III and IV.
     This results in high potential energy in the LUTM-AL field.
    }
  \label{fig:lutl_lutm}
\end{figure}


SLICEMs can operate in one of three modes: LUT, distributed RAM, or SHIFT, as
discussed in ~\secRef{sec:clb_heterogeneity}. In the LUT slots of that SLICEM,
only instances that match the mode can be placed. In order to address this
constraint, we introduce two electrostatic fields, \textit{LUTL} and \textit{LUTM-AL}, into the
multi-electrostatic placement model. In the field setup, it should be possible
to prevent distributed RAM instances or SHIFT instances from being placed on
SLICEL sites, however, it should be possible to place LUT instances both on
SLICEL and SLICEM sites, as part of the field setup.

There is an example of how these two fields are set up in \figRef{fig:lutl_lutm}.
LUTL models the LUT resources that are provided by both SLICEM and SLICEL,
whereas LUTM-AL models the additional logic resources that are supplied by SLICEM but not by SLICEL.
In contrast to a distributed RAM or SHIFT instance, a LUT instance only
occupies resources within the LUTL field, while a distributed RAM or SHIFT
instance occupies resources both in the LUTL and LUTM-AL fields.

Using an example of a LUT instance and a SHIFT instance as an example,
this figure analyses four scenarios.
As with SHIFT instances, distributed RAM instances work in exactly the same way.
In order to indicate that a SLICEL does not contain any resources for LUTM-AL,
we set the initial density for LUTM-AL in a SLICEL to 1, indicating the SLICEL is occupied;
in contrast, the initial density for LUTL in a SLICEL is set to 0.
We set the initial density of a SLICEM to zero for both LUTL and LUTM-AL due to the fact that it contains both LUTL and LUTM-AL resources.
In Solution I, if the LUT instance is placed on a SLICEL and the SHIFT instance
is placed on a SLICEM, there will be no overflow of density in either of the
fields (a balanced density distribution can be achieved by inserting fillers
\cite{PLACE_TODAES2015_Lu}), which means that the potential energy will be
minimized.
In Solution II, the scenario is similar to the one in Solution I.
Solution III and IV, on the other hand, where the SHIFT instance is placed on a
SLICEL, produce a density overflow in the LUTM-AL field, which indicates that
there is a high potential energy for this field.
As long as the optimizer minimizes the potential energy, these solutions will be avoided.
As a result, these two elaborate fields are able to accommodate LUT and
distributed RAM/SHIFT to their compatible sites in an easy manner.

\subsection{Divergence-aware Preconditioning}
\label{sec:preconditioning}
It is important to precondition the gradient $\nabla \mathcal{L}^{(t)}$ for each iteration $t$ before
it is fed to the optimizer, where $\mathcal{L}$ is the Lagrangian problem defined in \eqRef{eq:opt_alm}.
A gradient is discussed only in the direction of $x$, and a gradient in the direction of $y$ is the same.
To make the Jacobi preconditioner $\mathcal{P}$ more efficient,
we approximate the second-order derivatives of wirelength and density according to the following formula.
\begin{subequations}
  \begin{align}
    \mathcal{P}^{W}_i & = \frac{\partial^2 \widetilde{\mathcal{T}}_{\omega}(\boldsymbol{x}, \boldsymbol{y})}{\partial x_i^2} \sim \sum\limits_{e \in E_i} \frac{w_e}{\vert e \vert -1}, \quad \forall i \in \mathcal{V}, \label{eq:preconditioner_wl} \\
    \mathcal{P}_i^{(t)}& \sim \bigmax{1}{\bigbracket{\mathcal{P}^{W}_i
    + \sum_{s \in S} \alpha_{s}^{(t)}\lambda_{s}^{(t)}\mathcal{A}^s_i}^{-1}}, \label{eq:preconditioner}
  \end{align}
\end{subequations}

In the above equation, $\mathcal{V}$ denotes the set of instances, $E_i$ denotes the nets
incident to instance $i \in \mathcal{V}$, $w_e \in \omega$ denotes the weight of net $e$,
and $\mathcal{P}^{W}$ denotes the second-order derivative of the wirelength term.
To optimize the model, we provide the optimizer with the preconditioned gradient $\hat{\nabla}\mathcal{L}^{(t)} = \nabla \mathcal{L}^{(t)} \odot \mathcal{P}^{(t)}$.

\begin{figure}[tb]
   \includegraphics[width=\columnwidth]{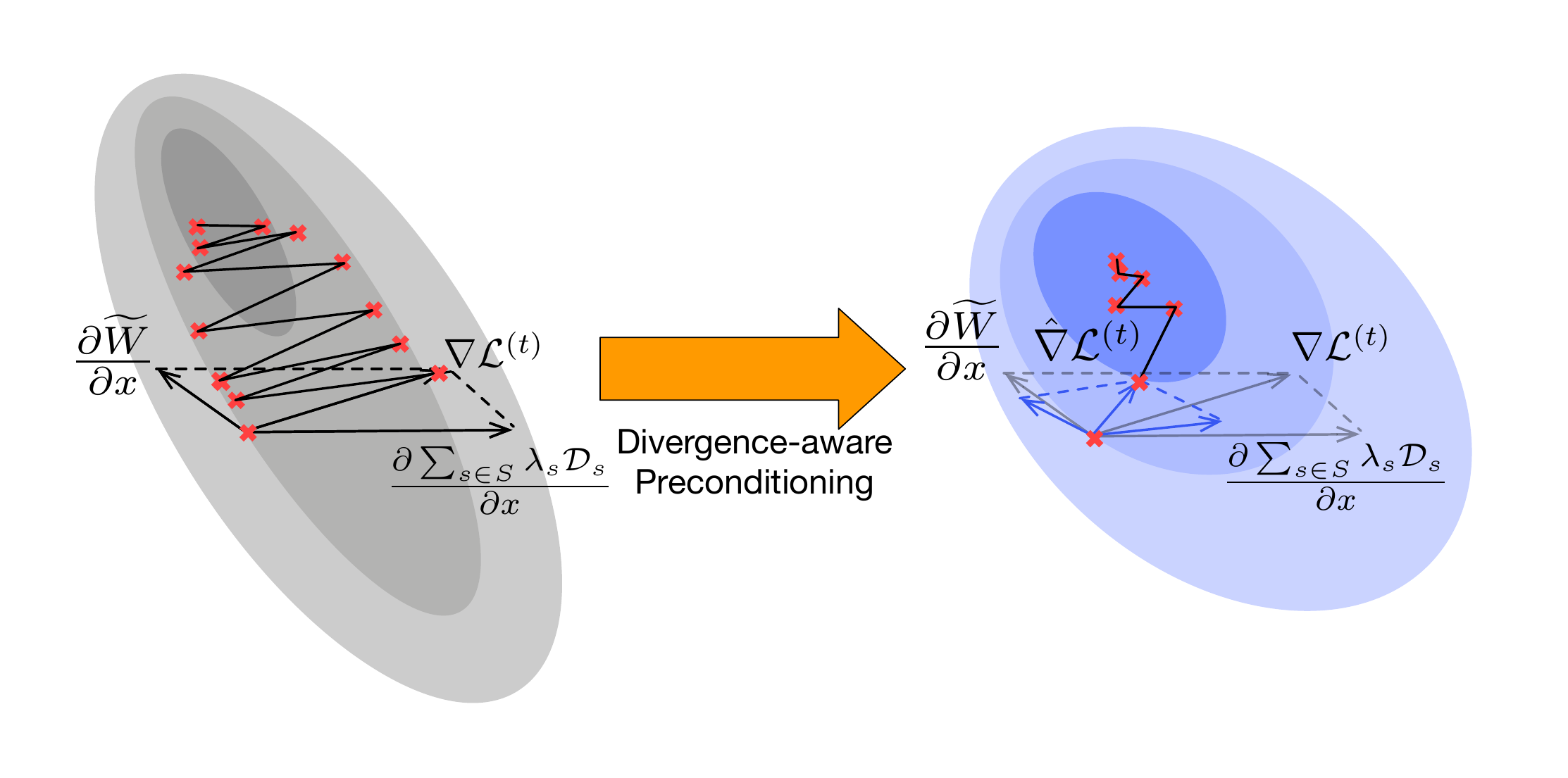}
   \caption{
    It is shown in the figure how preconditioning takes place at iteration
   $t$. A preconditioning technique translates objection surfaces into a more
   isotropic form, and this leads to a reduction in iterations and a
   stabilization of the optimization process as a result.
 }
   \label{fig:precond}
   \vspace{-.2in}
 \end{figure}

As shown in \figRef{fig:precond}, after the loss surfaces have been preconditioned, they are now more isotropic and therefore can be optimized more rapidly.
The intuition from the partial derivative itself is that
we would expect that for $\mathcal{P}^{W}_i$
instances with more pins or pins incident to larger net weights, they would move
slower than instances with fewer pins,
and for instances with larger $\sum_{s \in S} \alpha_{s}^{(t)}\lambda_{s}^{(t)}\mathcal{A}^s_i$ would also move slower.

It has been observed that if the ratio of the gradient norms from the density
term and the wirelength term becomes too large, the optimization can diverge
easily \cite{PLACE_TCAD2018_Cheng, PLACE_TCAD2021_Meng}.
Thus, we introduce an additional weighting vector $\vec{\alpha} \in \mathbb{R}^{\vert S \vert}$,
so that we can dynamically control the gradient norm ratio.
It is illustrated in \figRef{fig:precond} that some instances are dominated by
the density gradient, resulting in some instances moving too fast, and this
causes the instances to diverge from each other.
In most cases, this occurs during the second half of the placement iteration
when the density term starts to compete with the wirelength term by increasing
$\lambda$ to maintain its position in front of it.
In order to stabilize the optimization process, we need a new preconditioner.
For convenience, we define two auxiliary variables $\vec{\vartheta}^{(t)}$ and $\overline{\mathcal{\vec{P}}}^{W}$.
Taking the gradient norms of the density and wirelength terms to be equal, we can derive $\vec{\alpha}$ as follows.
\begin{subequations}
   \begin{align}
     \vartheta_s^{(t)} & = \bigmax{1}{
       \frac{\nabla \mathcal{D}_{s}}{\sum_{i \in \mathcal{V}^r_s} \vert \partial \widetilde{\mathcal{T}}_{\omega} / \partial x_i \vert}
       }, \quad \forall s \in S, \label{eq:vartheta} \\
     \overline{\mathcal{P}}^{W}_s & = \frac{\sum_{i \in \mathcal{V}^r_s}\mathcal{P}^{W}_i}{\vert \mathcal{V}^r_s \vert}, \quad \forall s \in S, \\
     \vec{\alpha}^{(t)} & = \vec{\vartheta}^{(t)} \odot \overline{\vec{\mathcal{P}}}^{W}, \label{eq:def_alpha}
   \end{align}
 \end{subequations}
$\mathcal{V}^r_s$ denotes the set of instances that have a demand in the field $s$.
$\sum_{i \in \mathcal{V}^r_s} \vert \partial \widetilde{\mathcal{T}}_{\omega} / \partial x_i \vert$ is the wirelength gradient norm summation of $\mathcal{V}^r_s$.
The weighting vector $\vec{\vartheta}^{(t)} \in \mathbb{R}^{\vert S \vert}$ measures the gradient norm ratio between the density term and the wirelength term,
and $\overline{\vec{\mathcal{P}}}^{W}$ denotes the average wirelength preconditioner for each field type.
The detailed derivations have been omitted for brevity.
Experiments will be conducted to further validate its effectiveness.

\subsection{Iterative Carry Chain Alignment Correction}
\label{sec:carry_alignment}

Using a carry chain alignment technique, we propose a method that can better align the carry chains without affecting the effectiveness of the analytical global placement algorithm.
We move the sequential CARRYs together at the end of each global placement iteration, align them based on their horizontal coordinates, and then move them into a column shape at the end of an iteration.
The CARRY instances in a chain will move together during the global placement iterations, which eases the legalization step, since the chains will almost align once the global placement process is completed.

\subsection{Clock Network Planning Algorithm}
\label{sec:cnp_algo}

There is a high degree of dissmoothness in the clock constraints in \secRef{sec:clock_constraint}.
The slightest movement within the clock region boundaries can result in an illegal clock configuration, which is detrimental to optimization.
In order to simplify the clock planning process, we decompose it into two stages.
Our first step is to find an \emph{instance-to-clock-region mapping} in the first stage.
This mapping ensures that all clock region constraints will be satisfied as long as all instances are located within the target clock region during the mapping.
The second step involves moving all instances to their target clock regions by
adding a penalty term to the placement objective, which is $\varGamma(\cdot)$.
Following the global placement, the half-column constraints are then dealt with
in the developed clock-aware direct legalization and detailed placement
algorithm \cite{PLACE_TCAD2019_Li_UTPLACEF_DL}.
As we move forward, we will explain these two stages in more detail.


\subsubsection{Instance-to-Clock-Region Mapping Generation}
\label{sec:cnp_mapping}

It is the goal of this step to generate mappings in a manner that ensures that clock constraints can be met with the minimum perturbation to the placement.
As discussed in \cite{PLACE_FPGA2019_Li}, we propose using
\emph{branch-and-bound method} to search through the solution space arising
from different instance-to-clock-region mappings, and find a feasible solution
with high quality within the solution space.
As part of the second stage, we enlist the assignment that has the lowest cost and uses it as a base.

\subsubsection{The Clock Penalty for Placement}
\label{sec:def_clock_penalty}

%
In contrast to \cite{PLACE_ICCAD2017_Pui_RippleFPGA, PLACE_TODAES2018_Li_UTPlaceF2}, which forces instances to move directly to their clock regions,
we introduce a bowl-like, smooth, and differentiable gravitational attraction term, which draws instances to the clock regions to which they have been mapped.
Let $lo^x_i$, $hi^x_i$, $lo^y_i$, and $hi^y_i$ be the left, right, bottom, and top boundary coordinates of the generated mapping result of instance $i$'.
We define the penalty term for instance $i$ as $\varGamma_i(\vec{x}_i, \vec{y}_i) = \varGamma_i(\vec{x}_i, \vec{y}_i)^x + \varGamma_i(\vec{x}_i, \vec{y}_i)^y$, where $ \varGamma_i(\vec{x}_i, \vec{y}_i)^x$ is defined as,
\begin{equation}
  \varGamma(\vec{x}_i, \vec{y}_i)^x = \left\{
    \begin{aligned}
      (\vec{x}_i - lo^x_i)^2,~& \vec{x}_i < lo^x_i, \\
      0, ~& lo^x_i \leq \vec{x}_i \leq hi^x_i, \\
      (\vec{x}_i - hi^x_i)^2, ~& hi^x_i < \vec{x}_i. \\
    \end{aligned}
  \right.
  \label{eq:clock_penalty}
\end{equation}
\begin{figure}[tb]
  \centering
  \includegraphics[width=0.6\columnwidth]{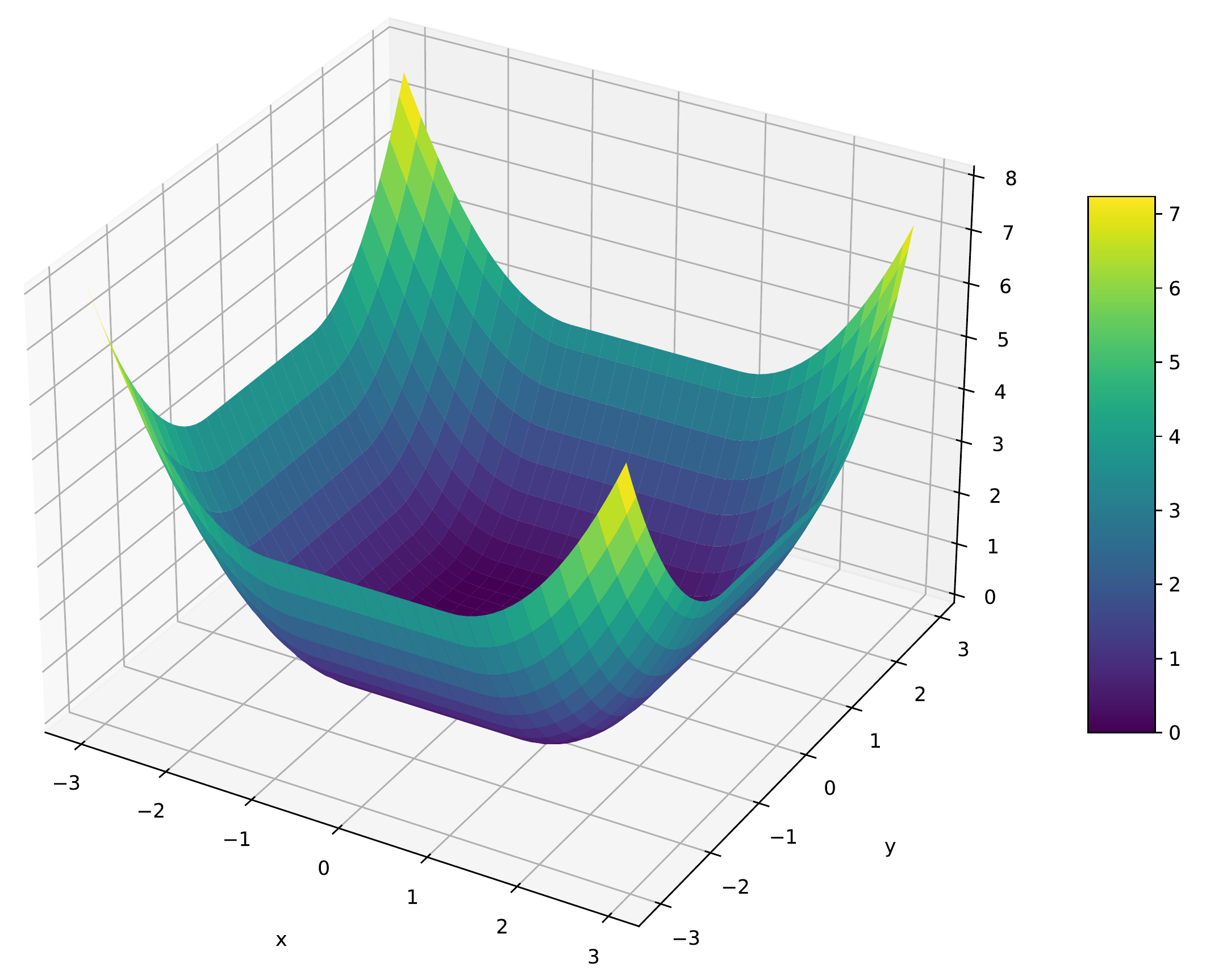}
  \caption{The Visualization of clock penalty function $\varGamma_i(\cdot)$ for a single instance.}
  \label{fig:clock_penalty}
\end{figure}
A visual representation of the clock penalty term can be found in \figRef{fig:clock_penalty}.
$\varGamma(\boldsymbol{x}, \boldsymbol{y})$ in \eqRef{eq:opt_relax} indicates that the sum of the clock penalty of all instances,
i.e., $\varGamma(\boldsymbol{x}, \boldsymbol{y})$ is equal to $\sum_{i \in \mathcal{V}}\varGamma_i(x_i, y_i)$.

Initially, the clock penalty multiplier $\eta$ is set to a value of 0.
As soon as we reset the clock penalty function $\varGamma(\cdot)$,
we update $\eta$ with the relative ratio between the gradient norms of the wirelength and the clock penalty in order to maintain the stability of the clock penalty function.
\begin{equation}
  \eta = \frac{\iota \norm{\nabla \widetilde{\mathcal{T}}_{\omega}}_1}{\norm{\nabla \varGamma}_1+ \varepsilon}.
\end{equation}
As we observe, only 1\% of the instances are out of their available clock regions right after they have been assigned the clock region, so most instances have no penalties as a result of this.
It is empirically established that $\iota$ is equal to $10^{-4}$ and $\varepsilon$ is equal to $10^{-2}$ as a method of balancing the gradient norm ratio.

\subsection{Timing-Criticality-based Weighting Method}
\label{sec:timing_opt}

\begin{figure*}[tb]
   \centering
   \includegraphics[width=\textwidth]{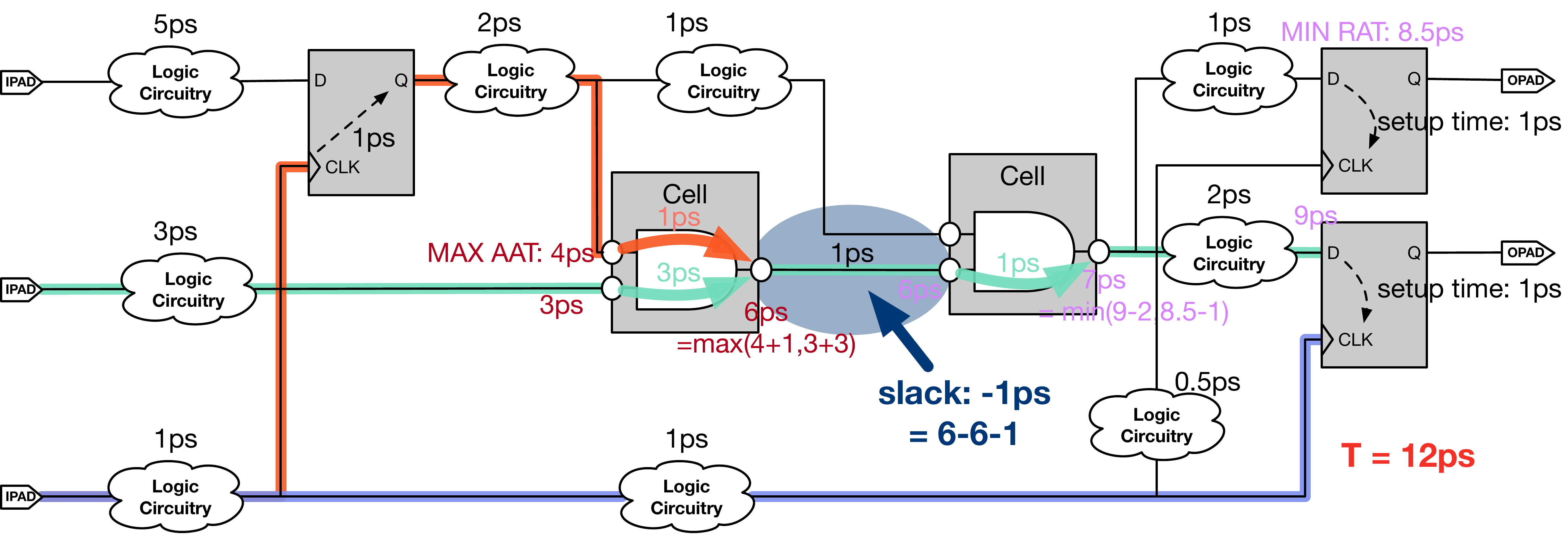}
   \caption{
     An illustration for the static timing analysis on FPGA. Support the clock period is $T$ and the signal arrival time at primary input ports has minor skew.
     The required arrival time at the sink of a timing path is the clock period minus the clock delay at the capture FFs and the setup time of FFs (1ps in our example),
     and the required arrival time of a timing endpoint is determined by the required arrival time of its fanout endpoints and the fanout edge delays.
     The actual arrival time of a timing endpoint is the maximal accumulated delay from the primary inputs.
     The slack of a timing edge is the difference between the required arrival time at its sink, and the summation of the actual arrival time at its source and its edge delay.
    }
   \label{fig:timing_analysis}
   \vspace{-.2in}
\end{figure*}

The timing performance objective $\widetilde{\mathcal{T}}_{\boldsymbol{\omega}}(\cdot)$ (see \eqRef{eq:opt_relax}) consists of two components:
i) the wirelength objective as a first-order approximation of WNS and TNS, and ii) the net criticality $\boldsymbol{\omega}$ that remedies the lack of timing information for first-order approximation.
\begin{equation}
  \widetilde{\mathcal{T}}_{\boldsymbol{\omega}}(\boldsymbol{x}, \boldsymbol{y}) = \sum\limits_{e \in E} \omega_e \cdot \widetilde{W}(e)
\end{equation}
$\omega_e \in \boldsymbol{\omega}$ measures the timing criticality of nets, i.e., 
whether any timing path through nets violates timing constraints, 
what is the degree of the violation, 
and how much the path delay can be reduced.
In this section we detail how $\widetilde{\mathcal{T}}(\cdot)$ functions.

\subsubsection{Static Timing Analysis}
\label{sec:static_timing_analysis}
Timing-driven placement leverages the static timing analysis (STA) to evaluate
the timing criticality of nets.
STA relies on (i) model of signal delays for nets and instances, and (ii) a timing analysis engine based on net and instance delays.

Delay models in FPGA are highly coupled with the pre-fabricated routing architecture and the behavior of the FPGA router.
In our experiment, we design the linear delay model, where the delay of the routing path between a source endpoint and a sink endpoint is calculated
by the Manhattan distance between them. 
Given the delay of the timing edges, the timing analysis engine determines which timing paths violate their timing constraints.
A timing path $\pi$ is a directed acyclic path for particular source and sink pairs (primary I/Os and I/Os of store elements).
The delay $t_\pi$ along a path $\pi$ is the sum of wire delays and cell delays,
and every path comes with a timing constraint $c_\pi$ defined via the \textit{actual arrival time (AAT)} and \textit{required arrival time (RAT)} for every driver pin and primary output.

The slack $s_\pi$ of a path $\pi$ is defined as $s_\pi = c_\pi - t_\pi$.
We mainly focus on the \textit{setup time} constraint, which is defined as the difference between the AAT and RAT.
A timing path $\pi$ violates the setup time constraints if the slack is negative.
The slack of a timing edge is the smallest path slack among the paths containing this edge.
To avoid enumerating all paths, we compute the slack from the actual arrival times and required arrival times at timing endpoints \cite{OpenTimer_ICCAD2015_Huang}.
\begin{subequations}
  \label{eq:att_rat}
  \begin{align}
    T_{ATT}(v_i) &= \max_{v_j \in fanin(v_i)} T_{ATT}(v_j) + e_{j,i} \\
    T_{ATT}(v_i) &= \min_{v_j \in fanout(v_i)} T_{RAT}(v_j) - e_{i,j}
  \end{align}
\end{subequations}
In \eqRef{eq:att_rat}, $v_i$ is an endpoint in the timing graph, and $e_{i,j}$ denotes the timing delays from endpoint $i$ to endpoint $j$, provided by the delay model.
The slack of timing edge $s_{i,j}$ connecting the source endpoint $v_i$ and the sink endpoint $v_j$ is
\begin{equation}
  s_{i,j} = T_{RAT}(v_j) - T_{ATT}(v_i) - e_{i,j},
\end{equation}
\figRef{fig:timing_analysis} illustrates the timing analysis process.

\subsubsection{Timing-driven Net Reweighting Scheme}
In the timing graph, nets have diverse effects on the timing slacks.
Those nets with higher timing criticality should be more sensitive to the timing closure.
To improve the timing in the analytical placement framework, we assign different net weights to different nets based on their timing criticality.
Let $s_e$ and $s_{wns}$ denote the timing slacks of net $e$ and the worst negative slack, respectively.
We define the timing criticality $c_e$ as follows,

\begin{equation}
  \label{eq:timing_crit}
  c_e = \frac{\min(0, s_e)}{\min(0, s_{wns})-T} \in [0, 1),
\end{equation}
where $T$ is the clock period.
%
When net $e$ is not on a path with timing violation, i.e. $s_e \geq 0$, the
timing criticality $c_e$ remains zero. Otherwise, the timing criticality equals the ratio $\frac{\vert s_e \vert}{\vert s_{wns} \vert+T}$.
The worse timing slack $\vert s_e \vert$ is, the larger timing criticality $c_e$ it will have.
The largest timing criticality falls upon the nets on the most critical path.

After evaluating the timing criticality, we compute the net weight $w_e \in \boldsymbol{\omega}$ as
\begin{subequations}
  \label{eq:timing_weight}
  \begin{align}
    \beta_e &= \alpha \cdot \max\left(1, \exp(c_e)\right) \\
    \omega_e' &\leftarrow \omega_e \cdot \beta_e \
  \end{align}
\end{subequations}
where $\omega_e'$ is the updated net weight, $\beta_e$ denotes the reweighting magnitude for net $e$, and $\alpha$ is a hyper-parameter that controls the weighting magnitude
\footnote{
In our experiments, $\alpha$ equals 1.
}.

\definecolor{tradewind}{HTML}{64B5AC}

\begin{figure}[tb]
    \centering
    \subfloat[]{\includegraphics[width=\columnwidth]{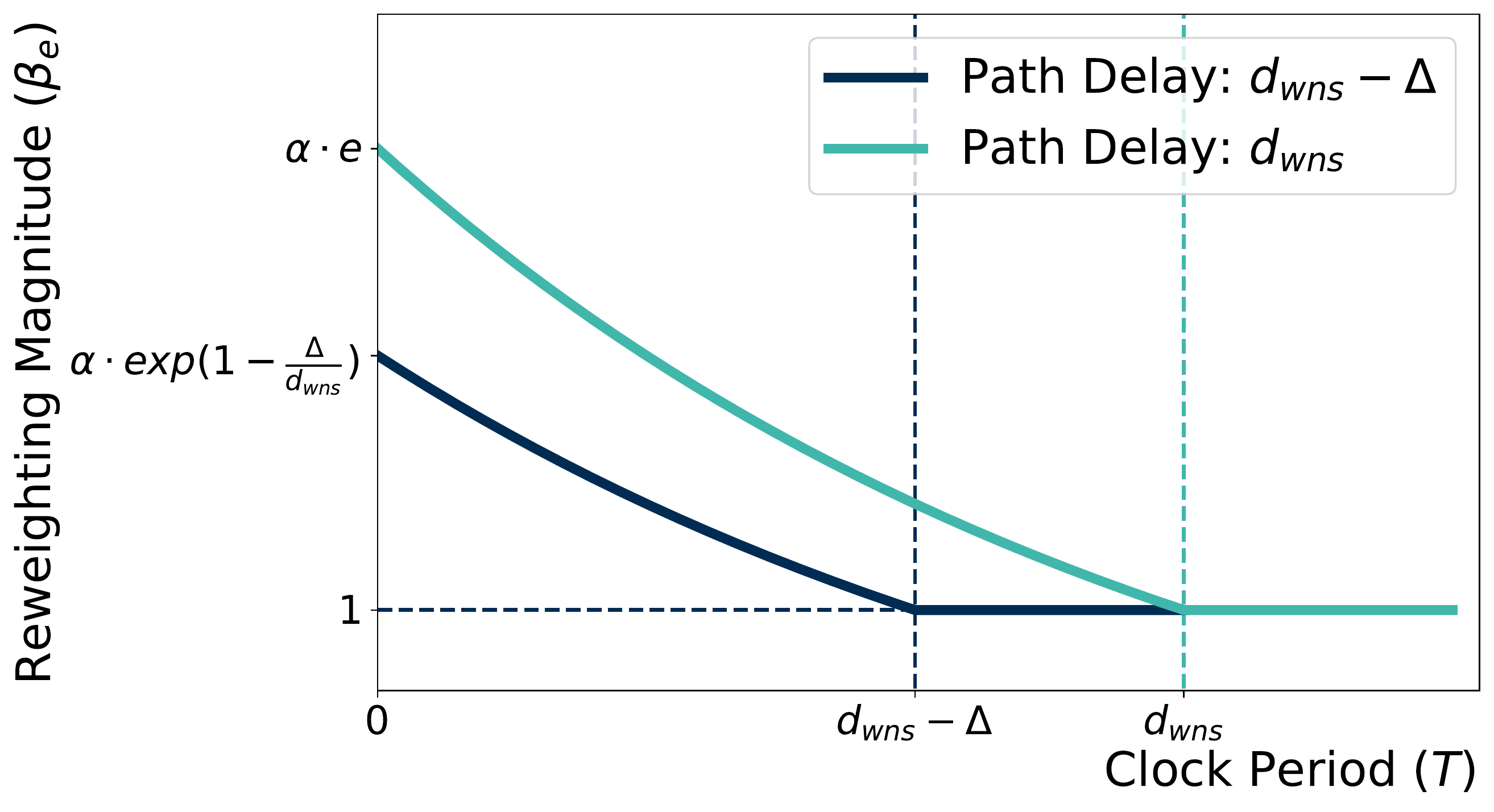}\label{fig:timing_w-t}} \hfill
    \subfloat[]{\includegraphics[width=\columnwidth]{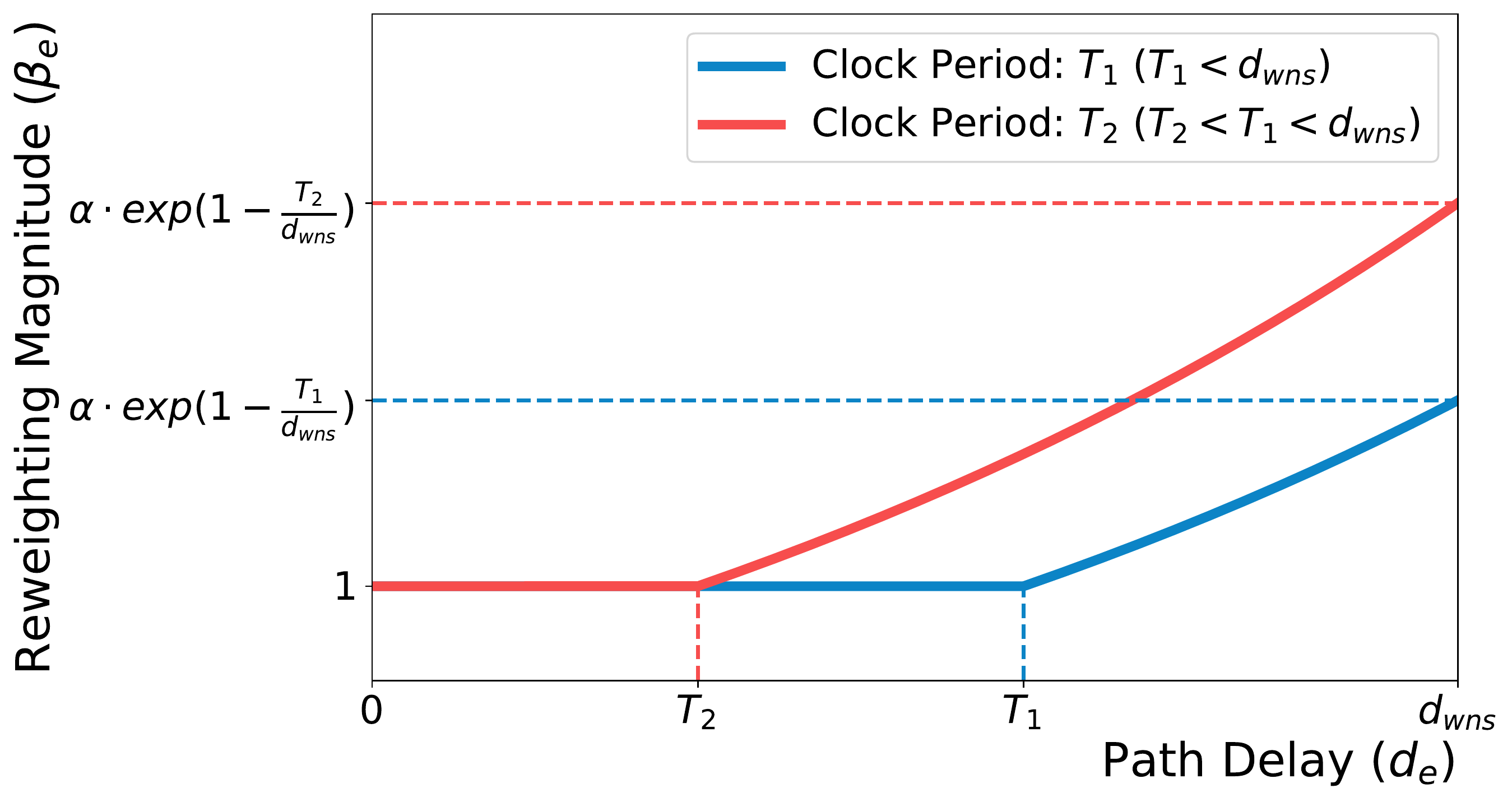}\label{fig:timing_w-d}}
    \caption{
      (a) This figure illustrates the relationship between the reweighting
      magnitude $\beta_e$ and the clock period $T$ given a certain path and its
      delay.
    In particular, the path with delay $d_{wns}$ represents the most critical
    path.
    This figure also depicts another path with delay $d_{wns}-\Delta$, where
    $\Delta$ measures how much the delay is smaller the worst-case path
    $d_{wns}$.
    (b) This figure shows the relationship between the reweighting magnitude
    $\beta_e$ and the path delay $d_e$ given a clock period.
    This figure plots two curves for two certain clock periods
    $T_1$ and $T_2$ ($T_2 < T_1 < d_{wns}$).
    %
    }
\end{figure}

\subsubsection{Effectiveness Analysis of the Rweighting Scheme}
\label{sec:tdp_effectiveness_analysis}
It is suggested that our reweighting scheme can control the reweighting scale according to the severity of the target timing constraints.
The reweighting magnitude $\beta_e$ of a net is determined by the largest magnitude of the timing paths that pass through it,
i.e., $\beta_e = \max\limits_{\pi \ni e} \beta_{\pi}$.
We then gave a brief analysis of how the timing path $\pi$ affects the re-weighting scheme in nets.

For the paths with maximal delay $d_{wns}$, the intuition is that the reweighting method should make no effect, i.e., $\beta_{wns}$ equals 1, when
it has no timing violations.
Otherwise, the smaller $T-d_{wns}$ is, the larger the reweighting magnitude
should be.
\figRef{fig:timing_w-t} show the relationship between the magnitude of the reweighting $\beta_{wns}$ and the clock period $T$ for a certain path and its delay. 
This figure demonstrates that a path will only gain a reweighting magnitude greater than one when the target clock period $T$ is less than its path delay.
The reweighting magnitude will grow faster with $T$ becoming smaller. 
\figRef{fig:timing_w-d} regards the path delay as a variable and shows how
the reweighting magnitude $\beta_{wns}$ reacts under a certain clock period.
The relationship curves align with our intuition that nets with a smaller timing delay
always gain a smaller reweighting magnitude.
No matter what the clock period is, the reweighting magnitude only takes effect
on those paths with timing violations. 
%

Nets on critical paths gain a larger weighting magnitude in this stage.
In the subsequent analytical placement step, the increased net weights
then help critical paths tilt more toward timing closure in the tradeoff
between timing and area.

\section{Experimental Results}
\label{sec:Results}


We implemented our GPU-accelerated placer in C++ and Python along with the open-source machine learning framework Pytorch for fast gradient back propagation\cite{PLACE_TCAD2020_Lin}.
We conduct experiments on a Ubuntu 18.04 LTS platform that consists of an Intel(R) Xeon(R) Silver 4214 CPU @ 2.20GHz (24 cores), one NVIDIA TITAN GPU,
and 251GB memory.
We demonstrate the effectiveness and efficiency of our proposed algorithm on both academic benchmarks \cite{BENCH_ISPD2017_PLACE} and industrial benchmarks
from the three most concerning aspects of routed wirelength (RWL), runtime (RT), and timing.

\subsection{Evaluation on Academic Benchmarks}

The statistics of the ISPD 2017 academic benchmark are summarized in \tabRef{tab:sota_comparison_ispd2017}.
The number of instances varies from 400K to 900K with 32--58 clock nets.
We do not evaluate the timing performance because on this benchmark we have no access to the timing information of the device.
We compare the routed wirelength reported by patched Xilinx Vivado v2016.4 and placement runtime with four state-of-the-art placers,
\texttt{UTPlaceF 2.0} \cite{PLACE_TODAES2018_Li_UTPlaceF2},
\texttt{RippleFPGA} \cite{PLACE_ICCAD2017_Pui_RippleFPGA},
\texttt{UTPlaceF 2.X} \cite{PLACE_FPGA2019_Li},
and \texttt{NTUfplace} \cite{PLACE_TCAD2020_Chen}.
All the results of these placers are from their original placers.
We do not compare the results with \texttt{elfPlace} because its algorithm cannot handle clock constraints (see \tabRef{tab:FPGAPlacers}).

\begin{table*}[tb]
	\centering
	\caption{Routed Wirelength (×$10^3$) and Runtime (Seconds) Comparison on ISPD 2017 Benchmarks.}
	\label{tab:sota_comparison_ispd2017}
  \resizebox{\textwidth}{!}{
\begin{tabular}{|c|cc|cc|cc|cc|cc|cc|}
\hline
  \multirow{2}{*}{Design} & \multirow{2}{*}{\#LUT/\#FF/\#BRAM/\#DSP} & \multirow{2}{*}{\#Clock} & \multicolumn{2}{c|}{\texttt{UTPlaceF 2.0} \cite{PLACE_TODAES2018_Li_UTPlaceF2}} & \multicolumn{2}{c|}{\texttt{RippleFPGA} \cite{PLACE_ICCAD2017_Pui_RippleFPGA}} & \multicolumn{2}{c|}{\texttt{UTPlaceF 2.X} \cite{PLACE_FPGA2019_Li}} & \multicolumn{2}{c|}{\texttt{NTUfplace} \cite{PLACE_TCAD2020_Chen}} & \multicolumn{2}{c|}{Ours (GPU)} \\
                        &                                         &                          & RWL              & RT              & RWL             & RT             & RWL    & \multicolumn{1}{c|}{RT}   & RWL             & RT            & RWL              & RT            \\ \hline \hline
\texttt{CLK-FGPA01}              & 211K/324K/164/75                        & 32                       & 2208            & 532             & 2011           & 288            & 2092  & \multicolumn{1}{c|}{180}  & 2039           & 698           & 1868           & 136           \\
\texttt{CLK-FGPA02}              & 230K/280K/236/112                       & 35                       & 2279            & 513             & 2168           & 266            & 2194  & \multicolumn{1}{c|}{179}  & 2149           & 710           & 2011           & 130           \\
\texttt{CLK-FGPA03}              & 410K/481K/850/395                       & 57                       & 5353            & 1039            & 5265           & 583            & 5109  & \multicolumn{1}{c|}{343}  & 4901           & 1704          & 4755           & 215           \\
\texttt{CLK-FGPA04}              & 309K/372K/467/224                       & 44                       & 3698            & 711             & 3607           & 380            & 3600  & \multicolumn{1}{c|}{242}  & 3614           & 1148          & 3338           & 162           \\
\texttt{CLK-FGPA05}              & 393K/469K/798/150                       & 56                       & 4692            & 939             & 4660           & 569            & 4556  & \multicolumn{1}{c|}{323}  & 4417           & 1540          & 4154           & 208           \\
\texttt{CLK-FGPA06}              & 425K/511K/872/420                       & 58                       & 5589            & 1066            & 5737           & 591            & 5432  & \multicolumn{1}{c|}{346}  & 5122           & 2210          & 4918           & 229           \\
\texttt{CLK-FGPA07}              & 254K/309K/313/149                       & 38                       & 2444            & 845             & 2326           & 304            & 2324  & \multicolumn{1}{c|}{201}  & 2320           & 795           & 2145           & 141           \\
\texttt{CLK-FGPA08}              & 212K/257K/161/75                        & 32                       & 1886            & 529             & 1778           & 247            & 1807  & \multicolumn{1}{c|}{169}  & 1803           & 588           & 1648           & 120           \\
\texttt{CLK-FGPA09}              & 231K/358K/236/112                       & 35                       & 2601            & 842             & 2530           & 327            & 2507  & \multicolumn{1}{c|}{197}  & 2436           & 717           & 2248           & 144           \\
\texttt{CLK-FGPA10}              & 327K/506K/542/255                       & 47                       & 4464            & 974             & 4496           & 512            & 4229  & \multicolumn{1}{c|}{286}  & 4339           & 1597          & 3839           & 200           \\
\texttt{CLK-FGPA11}              & 300K/468K/454/224                       & 44                       & 4183            & 1068            & 4190           & 455            & 3936  & \multicolumn{1}{c|}{265}  & 3964           & 1618          & 3626           & 183           \\
\texttt{CLK-FGPA12}              & 277K/430K/389/187                       & 41                       & 3369            & 774             & 3388           & 409            & 3236  & \multicolumn{1}{c|}{247}  & 3179           & 849           & 2938           & 168           \\
\texttt{CLK-FGPA13}              & 339K/405K/570/262                       & 47                       & 3816            & 1172            & 3833           & 441            & 3723  & \multicolumn{1}{c|}{270}  & 3680           & 985           & 3404           & 181           \\ 
  \hline \hline
Ratio                   &                                         &                          & 1.142           & 4.943           & 1.117          & 2.379          & 1.096 & 1.453                     & 1.079          & 6.575         & 1.000           & 1.000         \\ \hline
\end{tabular}%
  }
  \end{table*}%

\begin{table*}[tb]
  \centering
  \caption{Routed Wirelength ($\times 10^3$), WNS($\times 10^3$ps), TNS ($\times 10^5$ps) and Runtime (Seconds) Comparison between Conference Version and Our Algorithm on Industry Benchmarks.}
  \label{tab:industry_timing_results}
  \resizebox{\textwidth}{!}{
    \begin{tabular}{|c|cccc|cccc|cccc|}
      \hline
      \multirow{2}[2]{*}{Design} & \multirow{2}[2]{*}{\#LUT/\#FF/\#BRAM/\#DSP} & \multicolumn{1}{c}{\multirow{2}[2]{*}{\#Distributed \newline{}RAM+\#SHIFT}} & \multirow{2}[2]{*}{\#CARRY} & \multicolumn{1}{c|}{\multirow{2}[2]{*}{Clock\newline{}Period}} & \multicolumn{4}{c|}{Conference Version \cite{PLACE_DAC22_Mai}} & \multicolumn{4}{c|}{Ours (GPU)} \\
          &       &       &       &       & RWL   & WNS   & TNS   & RT    & RWL   & WNS   & TNS   & RT \\
      \hline \hline
      \texttt{IND01} & 17K/11K/0/13 & 9     & 2K    & 5     & 90    & -3.941 & -2.422 & 45    & 90    & -1.751 & -1.323 & 70 \\
      \texttt{IND02} & 11K/10K/0/24 & 6     & 335   & 2     & 102   & -4.240 & -19.733 & 54    & 117   & -2.938 & -18.148 & 76 \\
      \texttt{IND03} & 109K/12K/0/0 & 0     & 0     & 3     & 1028  & -2.666 & -18.467 & 59    & 1031  & -2.764 & -16.836 & 144 \\
      \texttt{IND04} & 29K/17K/0/16 & 218   & 1K    & 5     & 279   & -5.399 & -27.915 & 83    & 283   & -6.382 & -21.112 & 88 \\
      \texttt{IND05} & 64K/191K/64/928 & 29K   & 4K    & 10    & 2305  & -10.306 & -3.009 & 97    & 2312  & -4.558 & -2.354 & 218 \\
      \texttt{IND06} & 112K/65K/21/0 & 0     & 4K    & 15    & 1585  & -10.987 & -106.384 & 84    & 1585  & -6.502 & -51.922 & 193 \\
      \texttt{IND07} & 40K/156K/89/768 & 26K   & 3K    & 4     & 1498  & -6.302 & -20.585 & 83    & 1505  & -6.039 & -21.030 & 265 \\
      \hline \hline
    Ratio &       &       &       &       & 1.000  & 1.000  & 1.000  & 1.000  & 1.025  & 0.764  & 0.775  & 2.029 \\
      \hline
    \end{tabular}%
    }
\end{table*}%

\begin{table*}[htbp]
  \centering
  \caption{Routed Wirelength($\times 10^3$), WNS ($\times 10^3$ ps), TNS ($\times 10^5$ ps), and Runtime (Seconds) Comparison with Different Techniques on Industry Benchmarks.}
  \label{tab:industry_wl_results}%
  \resizebox{\textwidth}{!}{
    \begin{tabular}{|c|cccc|cccc|cccc|cccc|}
    \hline
    \multirow{3}[2]{*}{Design} & \multicolumn{4}{c|}{\multirow{2}[1]{*}{w/o precond or chain align\newline{}(GPU)}} & \multicolumn{4}{c|}{\multirow{2}[1]{*}{w/o precond\newline{}(GPU)}} & \multicolumn{4}{c|}{\multirow{2}[1]{*}{w/o chain align\newline{}(GPU)}} & \multicolumn{4}{c|}{\multirow{2}[1]{*}{Ours\newline{}(GPU)}} \\
          & \multicolumn{4}{c|}{}         & \multicolumn{4}{c|}{}         & \multicolumn{4}{c|}{}         & \multicolumn{4}{c|}{} \\
          & RWL   & WNS   & TNS   & RT    & RWL   & WNS   & TNS   & RT    & RWL   & WNS   & TNS   & RT    & RWL   & WNS   & TNS   & RT \\
    \hline \hline
      \texttt{IND01} & 103   & -3.491 & -2.849 & 53    & 94    & -3.041 & -1.826 & 66    & 108   & -2.581 & -143.009 & 74    & 90    & -1.751 & -1.323 & 70 \\
      \texttt{IND02} & 124   & -4.580 & -28.426 & 148   & 180   & -4.827 & -43.930 & 64    & 118   & -6.449 & -6029.390 & 73    & 117   & -2.938 & -18.148 & 76 \\
      \texttt{IND03} & 1021  & -3.080 & -17.301 & 123   & 1021  & -3.080 & -17.301 & 126   & 1030  & -2.604 & -1776.630 & 125   & 1031  & -2.764 & -16.836 & 144 \\
      \texttt{IND04} & 377   & -9.384 & -28.074 & 88    & 392   & -8.808 & -31.798 & 90    & 290   & -4.288 & -610.918 & 76    & 283   & -6.382 & -21.112 & 88 \\
      \texttt{IND05} & 2290  & -4.502 & -3.702 & 137   & diverge & diverge & diverge & diverge & 2260  & -4.870 & -238.678 & 153   & 2312  & -4.558 & -2.354 & 218 \\
      \texttt{IND06} & 1558  & -50.647 & -105.558 & 109   & 1576  & -11.450 & -113.733 & 107   & 1580  & -15.977 & -5320.100 & 114   & 1585  & -6.502 & -51.922 & 193 \\
      \texttt{IND07} & 1547  & -9.470 & -31.747 & 186   & 1393  & -8.145 & -20.816 & 155   & diverge & diverge & diverge & diverge & 1505  & -6.039 & -21.030 & 265 \\
    \hline\hline
    Ratio & 1.076 & 2.355 & 1.599 & 0.922 & 1.147 & 1.497 & 1.586 & 0.801 & 1.034 & 1.468 & 1.298 & 0.837 & 1.000 & 1.000 & 1.000 & 1.000 \\
    \hline
    \end{tabular}%
  }
\end{table*}%

The experimental results show that our placement algorithm consistently achieves better routed wirelengths than other placers.
Specifically, our placer achieves 14.2\% smaller routed wirelength than \texttt{UTPlaceF 2.0},
11.7\% smaller than \texttt{RippleFPGA},
9.6\% smaller than \texttt{UTPlaceF 2.X},
and 7.9\% smaller than \texttt{NTUfplace} on average, respectively.
For some benchmarks, like \texttt{CLK-FPGA06}, which is a large design with about 925K cells and 58 clock nets,
our routed wirelength is even 13.6\%, 6.7\%, 10.5\% and 4.1\% better than that of baseline placers, respectively.
It needs to be mentioned that \texttt{UTPlaceF 2.X} is a follow-up work to \texttt{UTPlaceF 2.0}, which relaxes the clock region bounding box constraints to clock tree constraints.
It allows for a larger solution space for clock routing feasibility and thus should yield better results.
However, even in this unfair comparison, our routed wirelength is still 9.6\% better than \texttt{UTPlaceF 2.X}, exhibiting the efficacy of our algorithm.
Besides, our GPU-accelerated placer is the fastest one with 4.94$\times$, 2.38$\times$, 1.45$\times$, and 6.58$\times$ speedup over other placers, respectively.
These experiments demonstrate the effectiveness and efficiency of our proposed algorithms.

\subsection{Evaluation on Industry Benchmarks}
We further evaluate our placer on industrial benchmarks which consist of a
comprehensive instance set and an industrial FPGA architecture from real-world
industry scenarios, including SHIFT, distributed RAM, and CARRY (see
\tabRef{tab:industry_timing_results}).
Most previous FPGA placers \cite{ TCAD18_RippleFPGA_Chen, PLACE_ICCAD2016_Ryan,
PLACE_TCAD2018_Li, PLACE_TCAD2021_Meng, TODAES18_GPlace3_Abuowaimer,
PLACE_ICCAD2017_Pui_RippleFPGA, PLACE_TODAES2018_Li_UTPlaceF2,
PLACE_FPGA2019_Li, PLACE_TCAD2020_Chen, PLACE_DATE2021_Lin } cannot fully
handle such an instance set with SLICEL-SLICEM heterogeneity.
To better validate the effectiveness of our placer, we leverage a high-quality
FPGA router to evaluate the placement algorithms more precisely \cite{ROUTE_ASPDAC2023_Wang}.

From \tabRef{tab:industry_timing_results}, we can see the comparison between the conference version \cite{PLACE_DAC22_Mai} and our placer.
Our placer can achieve 23.6\% better WNS and 22.5\% better TNS, respectively, with minor routed wirelength degeneration, exhibiting the effectiveness of our placer.
In some benchmarks, such as \texttt{IND06}, which is one of the most congestion benchmarks,
our WNS and TNS are 40.8\%  and 51.2\% better than the baseline, respectively.
These experiments demonstrate that our placer can effectively optimize timing even on congested benchmarks.
The experiments also show that our placer requires more time  to converge on industrial benchmarks. 
This is because optimizing timing requires additional iterations, which needs further optimization in the future.

\subsection{Ablation Study for Optimization Techniques}
To better understand the performance of our placer,
we perform an ablation study and validate the effectiveness of our proposed methods by disabling optimization techniques as follows (see \tabRef{tab:industry_wl_results}).
\begin{itemize}
\item Disable the iterative carry chain alignment correction in global placement and only align chains once in legalization.
\item Disable the dynamic preconditioner and use the default preconditioner in \cite{PLACE_TCAD2021_Meng, PLACE_TCAD2018_Cheng}.
\item Disable both techniques as the baseline.
\end{itemize}
We can see from \tabRef{tab:industry_wl_results} that both the iterative carry chain alignment correction technique
and dynamically-adjusted precondition technique helps stabilizes the global placement convergence and enables better
placement solution at convergence.
Without either of these two techniques, the placer fails to converge on one design.
The dynamically-adjusted preconditioner helps achieve 14.7\% better routed wirelength, 49.7\% better WNS, and 58.6\% better TNS
compared with the default preconditioner \cite{PLACE_TCAD2021_Meng, PLACE_TCAD2018_Cheng}.
Moreover, the iterative carry chain alignment correction helps optimize the routed wirelength, WNS,
and TNS by 3.4\%, 46.8\%, and 29.8\% with minor runtime overhead.
These experiments validate the effectiveness and efficacy of the proposed carry chain alignment and preconditioning technique.

\subsection{Runtime Breakdown}

\begin{figure}[tb]
    \centering
    \includegraphics[width=0.8\linewidth]{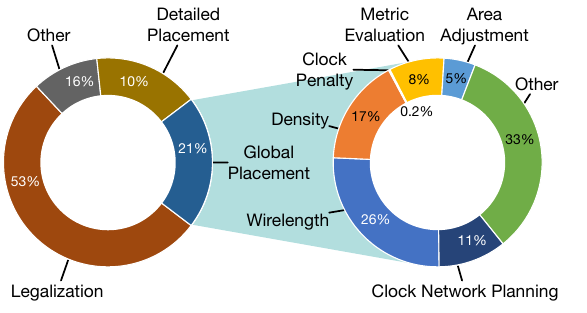}
    \caption{Runtime breakdown on \texttt{CLK-FPGA13}.
    Similar distributions are observed on other benchmarks.
    }
    \label{fig:runtime_breakdown}
    \vspace{-.25in}
\end{figure}

To analyze the time consumption of our placer, we further exhibit the runtime breakdown of our algorithm on one of the largest benchmarks as shown in \figRef{fig:runtime_breakdown}.
With the help of GPU acceleration, global placement is no longer the most time-consuming part.
The core forward and backward propagation of the global placement, i.e., the computation of wirelength and electrostatic density, as well as their gradients,
only take up 26\% and 17\% of the global placement runtime, respectively.
With the great reduction in global placement runtime,
legalization becomes the new runtime bottleneck, taking 53\% of the total placement time.
Meanwhile, other miscellaneous parts, including IOs, parsing, database establishment, etc., take up also the
same time as that of global placement.

\section{Conclusion}
\label{sec:Conclusion}
In this paper, we present a heterogeneous FPGA placement algorithm that
considers the heterogeneity of SLICELs and SLICEMs, as well as  timing
closure and clock feasibility.
A new electrostatic formulation and a nested Lagrangian paradigm have been
proposed to achieve uniform optimization of wirelength, routability, timing, and
clock feasibility for heterogeneous instance types, including LUT, FF, BRAM,
DSP, distributed RAM, SHIFT, and CARRY.
Additionally, we propose a dynamically adjusted preconditioner, a timing-driven net-weighting scheme, and a smooth clock penalization technique in order to ensure that the placement is convergent to high-quality solutions.
On the ISPD 2017 contest benchmark, experiments have revealed that our placer
can achieve 14.2\%, 11.7\%, 9.6\%, and 7.9\% better routed wirelengths compared
to the state-of-the-art placers, \texttt{UTPlaceF 2.0},
\texttt{RippleFPGA}, \texttt{UTPlaceF 2.X}, and \texttt{NTUfplace},
respectively, at 1.5-6$\times$ speedup leveraging GPU acceleration.
We conducted experiments on industrial benchmarks to prove that our algorithms
are capable of achieving 23.6\% better WNS and 22.5\% better TNS with about 2\% increase in routed wirelength.

%

{
\bibliographystyle{IEEEtran}
\bibliography{./ref/merged}

\begin{thebibliography}{10}
\providecommand{\url}[1]{#1}
\csname url@samestyle\endcsname
\providecommand{\newblock}{\relax}
\providecommand{\bibinfo}[2]{#2}
\providecommand{\BIBentrySTDinterwordspacing}{\spaceskip=0pt\relax}
\providecommand{\BIBentryALTinterwordstretchfactor}{4}
\providecommand{\BIBentryALTinterwordspacing}{\spaceskip=\fontdimen2\font plus
\BIBentryALTinterwordstretchfactor\fontdimen3\font minus
  \fontdimen4\font\relax}
\providecommand{\BIBforeignlanguage}[2]{{%
\expandafter\ifx\csname l@#1\endcsname\relax
\typeout{** WARNING: IEEEtran.bst: No hyphenation pattern has been}%
\typeout{** loaded for the language `#1'. Using the pattern for}%
\typeout{** the default language instead.}%
\else
\language=\csname l@#1\endcsname
\fi
#2}}
\providecommand{\BIBdecl}{\relax}
\BIBdecl

\bibitem{PLACE_2008_Lee}
S.-J. Lee and K.~Raahemifar, ``Fpga placement optimization methodology
  survey,'' in \emph{Canadian Conference on Electrical and Computer Engineering
  (CCECE)}.\hskip 1em plus 0.5em minus 0.4em\relax IEEE, 2008, pp.
  001\,981--001\,986.

\bibitem{PLACE_PIEEE2015_Markov}
I.~L. Markov, J.~Hu, and M.-C. Kim, ``Progress and challenges in {VLSI}
  placement research,'' \emph{Proceedings of the IEEE}, vol. 103, no.~11, pp.
  1985--2003, 2015.

\bibitem{PLACE_TCAD2005_Maidee}
P.~Maidee, C.~Ababei, and K.~Bazargan, ``Timing-driven partitioning-based
  placement for island style fpgas,'' \emph{IEEE TCAD}, vol.~24, no.~3, pp.
  395--406, 2005.

\bibitem{PLACE_DAC2015_ShengYen}
S.~Chen and Y.~Chang, ``Routing-architecture-aware analytical placement for
  heterogeneous fpgas,'' in \emph{Proc.~DAC}.\hskip 1em plus 0.5em minus
  0.4em\relax {ACM}, 2015, pp. 27:1--27:6.

\bibitem{PLACE_TCAD2018_Li}
W.~Li, S.~Dhar, and D.~Z. Pan, ``Utplacef: {A} routability-driven {FPGA} placer
  with physical and congestion aware packing,'' \emph{IEEE TCAD}, vol.~37,
  no.~4, pp. 869--882, 2018.

\bibitem{PLACE_ICCAD2016_Pui_RippleFPGA}
C.~Pui, G.~Chen, W.~Chow, K.~Lam, J.~Kuang, P.~Tu, H.~Zhang, E.~F.~Y. Young,
  and B.~Yu, ``Ripplefpga: a routability-driven placement for large-scale
  heterogeneous fpgas,'' in \emph{Proc.~ICCAD}, 2016, p.~67.

\bibitem{PLACE_ICCAD2016_Ryan}
R.~Pattison, Z.~Abuowaimer, S.~Areibi, G.~Gr{\'{e}}wal, and A.~Vannelli,
  ``Gplace: a congestion-aware placement tool for ultrascale fpgas,'' in
  \emph{Proc.~ICCAD}, 2016, p.~68.

\bibitem{PLACE_TCAD2019_Li_UTPLACEF_DL}
W.~Li and D.~Z. Pan, ``A new paradigm for {FPGA} placement without explicit
  packing,'' \emph{IEEE TCAD}, vol.~38, no.~11, pp. 2113--2126, 2019.

\bibitem{PLACE_TODAES2018_Li_UTPlaceF2}
W.~Li, Y.~Lin, M.~Li, S.~Dhar, and D.~Z. Pan, ``Utplacef 2.0: {A}
  high-performance clock-aware {FPGA} placement engine,'' \emph{ACM TODAES},
  vol.~23, no.~4, pp. 42:1--42:23, 2018.

\bibitem{PLACE_FPGA2019_Li}
W.~Li, M.~E. Dehkordi, S.~Yang, and D.~Z. Pan, ``Simultaneous placement and
  clock tree construction for modern fpgas,'' in \emph{Proc.~FPGA}, 2019, pp.
  132--141.

\bibitem{PLACE_ICCAD2017_Pui_RippleFPGA}
C.~Pui, G.~Chen, Y.~Ma, E.~F.~Y. Young, and B.~Yu, ``Clock-aware ultrascale
  {FPGA} placement with machine learning routability prediction: (invited
  paper),'' in \emph{Proc.~ICCAD}.\hskip 1em plus 0.5em minus 0.4em\relax
  {IEEE}, 2017, pp. 929--936.

\bibitem{PLACE_ICCAD21_Liang_AMFPlacer}
T.~Liang, G.~Chen, J.~Zhao, L.~Feng, S.~Sinha, and W.~Zhang, ``Amf-placer:
  High-performance analytical mixed-size placer for fpga,'' in
  \emph{Proc.~ICCAD}, 2021, pp. 1--6.

\bibitem{PLACE_TCAD2021_Meng}
Y.~Meng, W.~Li, Y.~Lin, and D.~Z. Pan, ``elfplace: Electrostatics-based
  placement for large-scale heterogeneous fpgas,'' \emph{IEEE TCAD}, 2021.

\bibitem{PLACE_ICCAD2017_Kuo_NTUfplace}
Y.~Kuo, C.~Huang, S.~Chen, C.~Chiang, Y.~Chang, and S.~Kuo, ``Clock-aware
  placement for large-scale heterogeneous fpgas,'' in \emph{Proc.~ICCAD}, 2017,
  pp. 519--526.

\bibitem{PLACE_TCAD2020_Chen}
J.~Chen, Z.~Lin, Y.~Kuo, C.~Huang, Y.~Chang, S.~Chen, C.~Chiang, and S.~Kuo,
  ``Clock-aware placement for large-scale heterogeneous fpgas,'' \emph{IEEE
  TCAD}, vol.~39, no.~12, pp. 5042--5055, 2020.

\bibitem{BENCH_ISPD2016_PLACE}
S.~Yang, A.~Gayasen, C.~Mulpuri, S.~Reddy, and R.~Aggarwal,
  ``Routability-driven {FPGA} placement contest,'' in \emph{Proc.~ISPD}, 2016,
  pp. 139--143.

\bibitem{BENCH_ISPD2017_PLACE}
S.~Yang, C.~Mulpuri, S.~Reddy, M.~Kalase, S.~Dasasathyan, M.~E. Dehkordi,
  M.~Tom, and R.~Aggarwal, ``Clock-aware {FPGA} placement contest,'' in
  \emph{Proc.~ISPD}, 2017, pp. 159--164.

\bibitem{PLACE_ICCAD2013_Contest}
M.-C. Kim, N.~Viswanathan, Z.~Li, and C.~Alpert, ``{ICCAD-2013 CAD} contest in
  placement finishing and benchmark suite,'' in \emph{Proc.~ICCAD}, 2013, pp.
  268--270.

\bibitem{DI_ULTRASCALE_CLB}
\BIBentryALTinterwordspacing
Ultrascale architecture clb slices. [Online]. Available:
  \url{https://www.xilinx.com/support/documentation/user_guides/ug574-ultrascale-clb.pdf}
\BIBentrySTDinterwordspacing

\bibitem{PLACE_DAC2019_Martin}
T.~Martin, D.~Maarouf, Z.~Abuowaimer, A.~Alhyari, G.~Gr{\'{e}}wal, and
  S.~Areibi, ``A flat timing-driven placement flow for modern fpgas,'' in
  \emph{Proc.~DAC}.\hskip 1em plus 0.5em minus 0.4em\relax {ACM}, 2019, p.~4.

\bibitem{PLACE_DATE2021_Lin}
Z.~Lin, Y.~Xie, G.~Qian, J.~Chen, S.~Wang, J.~Yu, and Y.-W. Chang,
  ``Timing-driven placement for fpgas with heterogeneous architectures and
  clock constraints,'' in \emph{Proc.~DATE}.\hskip 1em plus 0.5em minus
  0.4em\relax IEEE, 2021, pp. 1564--1569.

\bibitem{PLACE_FPGA2000_Marquardt}
A.~Marquardt, V.~Betz, and J.~Rose, ``Timing-driven placement for fpgas,'' in
  \emph{Proc.~FPGA}.\hskip 1em plus 0.5em minus 0.4em\relax {ACM}, 2000, pp.
  203--213.

\bibitem{TCAD18_RippleFPGA_Chen}
G.~Chen, C.-W. Pui, W.-K. Chow, K.-C. Lam, J.~Kuang, E.~F. Young, and B.~Yu,
  ``{RippleFPGA: Routability-driven simultaneous packing and placement for
  modern FPGAs},'' \emph{IEEE TCAD}, vol.~37, no.~10, pp. 2022--2035, 2018.

\bibitem{TODAES18_GPlace3_Abuowaimer}
Z.~Abuowaimer, D.~Maarouf, T.~Martin, J.~Foxcroft, G.~Gr{\'e}wal, S.~Areibi,
  and A.~Vannelli, ``{GPlace3.0: Routability-driven analytic placer for
  UltraScale FPGA architectures},'' \emph{ACM TODAES}, vol.~23, no.~5, pp.
  66:1--66:33, 2018.

\bibitem{DI_ULTRASCALE}
\BIBentryALTinterwordspacing
Ultrascale architecture clocking resources. [Online]. Available:
  \url{https://www.xilinx.com/support/documentation/user_guides/ug572-ultrascale-clocking.pdf}
\BIBentrySTDinterwordspacing

\bibitem{PLACE_TODAES2015_Lu}
J.~Lu, P.~Chen, C.-C. Chang, L.~Sha, D.~J.-H. Huang, C.-C. Teng, and C.-K.
  Cheng, ``e{P}lace: Electrostatics-based placement using fast fourier
  transform and nesterov's method,'' \emph{ACM TODAES}, vol.~20, no.~2, p.~17,
  2015.

\bibitem{PLACE_ISPD2002_Kahng}
A.~B. Kahng, S.~Mantik, and I.~L. Markov, ``Min-max placement for large-scale
  timing optimization,'' in \emph{Proc.~ISPD}, 2002, pp. 143--148.

\bibitem{PLACE_TCAD2015_Lu}
J.~Lu, H.~Zhuang, P.~Chen, H.~Chang, C.-C. Chang, Y.-C. Wong, L.~Sha, D.~Huang,
  Y.~Luo, C.-C. Teng \emph{et~al.}, ``{ePlace-MS}: Electrostatics-based
  placement for mixed-size circuits,'' \emph{IEEE TCAD}, vol.~34, no.~5, pp.
  685--698, 2015.

\bibitem{andreani2008augmented}
R.~Andreani, E.~G. Birgin, J.~M. Mart{\'\i}nez, and M.~L. Schuverdt, ``On
  augmented lagrangian methods with general lower-level constraints,''
  \emph{SIAM Journal on Optimization}, vol.~18, no.~4, pp. 1286--1309, 2008.

\bibitem{PLACE_TCAD2018_Cheng}
C.-K. Cheng, A.~B. Kahng, I.~Kang, and L.~Wang, ``Replace: Advancing solution
  quality and routability validation in global placement,'' \emph{IEEE TCAD},
  2018.

\bibitem{OpenTimer_ICCAD2015_Huang}
T.~Huang and M.~D.~F. Wong, ``Opentimer: {A} high-performance timing analysis
  tool,'' in \emph{Proc.~ICCAD}.\hskip 1em plus 0.5em minus 0.4em\relax {IEEE},
  2015, pp. 895--902.

\bibitem{PLACE_TCAD2020_Lin}
Y.~Lin, Z.~Jiang, J.~Gu, W.~Li, S.~Dhar, H.~Ren, B.~Khailany, and D.~Z. Pan,
  ``{DREAMPlace}: Deep learning toolkit-enabled gpu acceleration for modern
  vlsi placement,'' \emph{IEEE TCAD}, June 2020.

\bibitem{PLACE_DAC22_Mai}
J.~Mai, Y.~Meng, Z.~Di, and Y.~Lin, ``Multi-electrostatic fpga placement
  considering slicel-slicem heterogeneity and clock feasibility,'' in
  \emph{Proc.~DAC}, 2022, pp. 649--654.

\bibitem{ROUTE_ASPDAC2023_Wang}
J.~Wang, J.~Mai, Z.~Di, and Y.~Lin, ``A robust fpga router with concurrent
  intra-clb rerouting,'' in \emph{Proc.~ASPDAC}, 2023, pp. 529--534.

\end{thebibliography}
}

\end{document}